\documentclass[lettersize,journal]{IEEEtran}
\usepackage{amsmath,amsfonts}
\usepackage{algorithmic}
\usepackage{algorithm}
\usepackage{array}
\usepackage[caption=false,font=normalsize,labelfont=sf,textfont=sf]{subfig}
\usepackage{textcomp}
\usepackage{stfloats}
\usepackage{url}
\usepackage{verbatim}
\usepackage{graphicx}
\usepackage[dvipsnames]{xcolor}
\usepackage{cite}
\DeclareGraphicsExtensions{.pdf,.png,.jpg}

\newcommand {\Define} {\stackrel {\Delta} {=}  }
\newcommand{\mya}{\mathrel{\overset{\makebox[0pt]{{\tiny(a)}}}{=}}}
\newcommand{\myb}{\mathrel{\overset{\makebox[0pt]{{\tiny(b)}}}{=}}}
\newcommand{\myc}{\mathrel{\overset{\makebox[0pt]{{\tiny(c)}}}{=}}}
\newcommand{\myd}{\mathrel{\overset{\makebox[0pt]{{\tiny(d)}}}{=}}}

\newtheorem{theorem}{Theorem}

\newtheorem{result}{Result}
\usepackage{cite}
\usepackage{comment}

\begin{document}
	\title{Zak-OTFS over CP-OFDM}
	\author{\IEEEauthorblockN{Saif Khan Mohammed, Saurabh Prakash, Muhammad Ubadah, Imran Ali Khan, Ronny Hadani, Shlomo Rakib, Shachar Kons, Yoav Hebron, Ananthanarayanan Chockalingam and Robert Calderbank}
		\IEEEauthorblockA{ \thanks{S. K. Mohammed, S. Prakash, M. Ubadah and I. A. Khan are with the Department of Electrical Engineering, Indian Institute of Technology Delhi, India (E-mail: saifkmohammed@gmail.com, eez188572@ee.iitd.ac.in, eez198356@ee.iitd.ac.in, eez188166@ee.iitd.ac.in).
S. K. Mohammed is also associated with the Bharti School of Telecom. Technology and Management (BSTTM), IIT Delhi, India. R. Hadani is with the Department of Mathematics, University of Texas at Austin, USA (email: hadani@math.utexas.edu). R. Hadani is also with Cohere Technologies Inc., Santa Clara, CA, USA.
S. Rakib, S. Kons and Y. Hebron are with Cohere Technologies Inc., Santa Clara, CA, USA (email: shlomo.rakib@cohere-tech.com, shachar.kons@cohere-tech.com, yoav.hebron@cohere-tech.com).
A. Chockalingam is with the Department of Electrical Communication Engineering, Indian Institute of Science Bangalore, India (email: achockal@iisc.ac.in).
R. Calderbank is with the Department of Electrical and Computer Engineering, Duke University, Durham, NC, 27708, USA (email: robert.calderbank@duke.edu).}    }
	}
	\maketitle
	
	\begin{abstract}
    Zak-Orthogonal Time Frequency Space (Zak-OTFS) modulation
    has been shown to achieve significantly better performance compared to the standardized Cyclic-Prefix Orthogonal Frequency Division Multiplexing (CP-OFDM), in high delay/Doppler spread scenarios envisaged in next generation communication systems. Zak-OTFS carriers are quasi-periodic pulses in the delay-Doppler (DD) domain, characterized by two parameters, (i) the pulse period along the delay axis (``delay period") (Doppler period is related to the delay period), and (ii) the pulse shaping filter.
    
    An important practical challenge is enabling support for Zak-OTFS modulation in existing CP-OFDM based modems.
    In this paper we show that Zak-OTFS modulation with
    pulse shaping constrained to sinc filtering (filter bandwidth equal to the communication bandwidth $B$) followed by time-windowing with a rectangular window of duration $(T + T_{cp})$ ($T$ is the symbol duration and $T_{cp}$ is the CP duration), can be implemented as a low-complexity precoder over standard CP-OFDM. We also show that the Zak-OTFS de-modulator with matched filtering constrained to sinc filtering (filter bandwidth $B$) followed by rectangular time windowing over duration $T$ can be implemented as a low-complexity post-processing of the CP-OFDM de-modulator output. This
    proposed \emph{``Zak-OTFS over CP-OFDM"} architecture enables us to harness the benefits of
    Zak-OTFS in existing network infrastructure. We also show that the proposed Zak-OTFS over CP-OFDM is a family of modulations, with CP-OFDM being a special case when the delay period takes its minimum possible value equal to the inverse bandwidth, i.e., Zak-OTFS over CP-OFDM with minimum delay period. Despite the fact that
    the pulse shaping filter parameter is constrained in Zak-OTFS over CP-OFDM, numerical simulations for high Doppler spread scenarios show that it achieves significantly better effective spectral efficiency (SE) when compared to CP-OFDM. Unconstrained Zak-OTFS (where we are free to choose the delay/Doppler period and transmit pulse shaping and matched receiver filters) is shown to achieve effective SE performance even better than Zak-OTFS over CP-OFDM.
    The proposed Zak-OTFS over CP-OFDM architecture can also be extended to
    filtered OFDM.
    \end{abstract} 
	

\normalsize
\section{Introduction}
Existing cellular wireless communication systems are based on the standardized Cyclic-Prefix (CP) - Orthogonal Frequency Division Multiplexing (OFDM) modulation
\cite{IMT2020, 3GPP}. Next generation communication systems are envisaged to
extend the capabilities of existing systems to further support ubiquitous communication, integrated sensing and communication (ISAC) and Artificial Intelligence (AI)
for wireless communication \cite{IMT2030}. Ubiquitous communication can be achieved through reliable and high throughput non-terrestrial networks (NTN) and their integration with terrestrial networks, for example Low Earth Orbit (LEO) satellite based NTN. Efficient design of ISAC systems requires good communication and sensing waveforms which coexist on the same physical resource creating minimal interference for each other \cite{ISAC}. It is envisaged that AI will be integrated into the core architecture of next generation wireless communication systems to build more efficient,
intelligent and autonomous networks \cite{AI6G}.

However, these new scenarios bring along their own challenges. LEO satellite channels are characterized by very high Doppler shifts which limit the throughput and reliability of OFDM based systems \cite{Nee2000, Wang2006}. Sinusoidal OFDM carriers are known to be sub-optimal sensing waveforms and therefore we need new waveforms which are good for both sensing as well as for communication \cite{OFDMISAC,FMCWradar}. AI can improve network performance through intelligent resource allocation, interference management, etc.
This however assumes an accurate and timely prediction of the propagation channel conditions.
With OFDM, the effective channel is both time and frequency selective which makes it difficult to accurately predict the channel. In the absence of accurate prediction, it would be challenging to achieve
the benefits of AI based network management.

Recently introduced Zak-Orthogonal Time Frequency Space (Zak-OTFS) modulation is known to exhibit several advantages over OFDM in doubly-spread/time-varying channels \cite{ZAKOTFS1, ZAKOTFS2, otfsbook}.
Zak-OTFS carriers are narrow quasi-periodic pulses in the delay-Doppler (DD) domain characterized by their period along the delay and Doppler axis. The interaction of these DD carriers with a doubly-spread channel is \emph{predictable}. That is,
the channel response to a particular DD carrier (e.g., a pilot carrier) can be used to accurately predict the channel response to any other carrier. This results in an almost stationary input-output (I/O) relation in the DD domain. Due to the predictable and almost stationary nature of the I/O relation, it can be estimated/acquired with very low overhead (a single DD pilot is sufficient). Joint equalization of all carriers results in throughput and reliability performance which is almost invariant to the channel delay and Doppler spread, as long as the crystallization condition is satisfied (i.e., the delay and Doppler period of the carriers is greater than the channel delay and Doppler spread respectively). On the other hand, OFDM has a non-predictable I/O relation in doubly-spread channels which is difficult to efficiently acquire/estimate, resulting in high channel estimation overhead when the channel delay/Doppler spread is high. Zak-OTFS is also a good radar/sensing waveform with performance better than standard Chirp/Linear frequency modulated waveforms \cite{Danishradar}.
Zak-OTFS can also enable AI based intelligent network management since the effective DD domain channel is almost stationary when compared to the TF domain channel experienced by OFDM. Zak-OTFS is therefore suited for
the new scenarios envisaged in next generation wireless communication systems.

There are however challenges with regards to practical implementation of Zak-OTFS which need to be addressed. Since Zak-OTFS achieves robustness to high delay/Doppler spread through joint equalization of all carriers, its equalization/detection complexity (cubic in the number of carriers) can be prohibitive for user terminals having low signal processing capability. This issue has however been addressed recently in
\cite{SaifFDEOTFS, HZhang}, where low-complexity Zak-OTFS equalization methods have been proposed having complexity which is only square in the number of carriers. Also, recently, a novel DD domain precoding of Zak-OTFS signals is proposed in \cite{SaifPrecZakOTFS} which results in an almost ideal precoded channel for which per-carrier receiver equalization
is almost optimal, thereby reducing the receiver equalization complexity to only linear in the number of carriers . OFDM is well known for its ability to multiplex users by
allocating non-overlapping time-frequency (TF) resources to different users. Recently, in \cite{SaifMU} it has been shown that it is possible to achieve the same for Zak-OTFS.

However, one of the most important challenge is to enable support for Zak-OTFS modulation in existing 4G/5G modems.
A Zak-OTFS carrier is a narrow pulse in the DD domain,
characterized by, (i) its period along the delay axis (``delay period"), and (ii) the pulse shaping filter to constrain the
bandwidth and time-duration of the Zak-OTFS modulated signal.
The Doppler period of the DD pulse carrier is related to its delay period. In this paper we show that Zak-OTFS modulation with pulse
shaping constrained to sinc filtering (filter bandwidth equal to the communication bandwidth) followed by time-windowing with a rectangular time-window limited to the CP-OFDM symbol duration (including CP), can be implemented as precoder followed by the standard CP-OFDM modulator. To be precise, in this proposed ``Zak-OTFS over CP-OFDM" architecture, the DD domain information symbols are precoded into frequency domain (FD) symbols through the Inverse Discrete Frequency Zak Transform (IDFZT), followed by the standard CP-OFDM modulator which converts these FD symbols into the continuous time transmit signal. The precoding operation has a low complexity of only ${\mathcal O}(K \log K)$ where $K$ is the number of CP-OFDM sub-carriers and can be implemented in software using standard Fast Fourier Transform (FFT) libraries.

We also show that the Zak-OTFS demodulator with matched filtering constrained to sinc filtering (filter bandwidth equal to the communication bandwidth) followed by rectangular time windowing (duration equal to that of the OFDM symbol without CP), can be implemented as a post-processing of the standard CP-OFDM demodulator output. To be precise, FD symbols received on the CP-OFDM sub-carriers are converted to DD domain through the Discrete Frequency Zak Transform (DFZT) post-processor. Discrete Frequency Zak Transform (DFZT) can be implemented in software using FFT libraries and has a ${\mathcal O}(K \log K)$ complexity. 

The main underlying novel idea behind the proposed ``Zak-OTFS over CP-OFDM" architecture is as follows. In a Zak-OTFS modulator, the DD domain symbols are directly converted to time-domain symbols using the Inverse Discrete Zak Transform (IDZT), followed by pulse shaping (filtering and time-windowing). IDZT can be expressed as a cascade of IDFZT followed by the Inverse Discrete Fourier Transform (IDFT). With pulse shaping constrained to sinc filtering followed by rectangular time-widowing, the signal processing from the input of the IDFT module to the continuous-time transmitter output signal is nothing but the standard CP-OFDM modulator. A similar thing is true for the receiver processing in a Zak-OTFS receiver where the Discrete Zak Transform (DZT) converts the received discrete-time samples to DD domain symbols. DZT can be expressed as a cascade of DFT followed by the Discrete Frequency Zak Transform (DFZT). When the matched filtering comprises of sinc filtering followed by rectangular windowing, the signal processing from the received continuous time signal to the output of the DFT is nothing but a standard CP-OFDM demodulator. 

The proposed Zak-OTFS over CP-OFDM architecture therefore enables Zak-OTFS to be practically implemented over existing CP-OFDM based 4G/5G modems without changing existing modem hardware. This flexibility
allows existing network infrastructure to harness the benefits of Zak-OTFS which would help them address the new requirements/scenarios of next generation communication systems.
However, the proposed Zak-OTFS over CP-OFDM architecture constrains the Zak-OTFS transmit pulse shaping filter and the receiver matched filter to be equivalent to time-domain pulse shaping and match filtering in CP-OFDM. Also, since the duration of the Zak-OTFS packet/frame should coincide with the CP-OFDM symbol duration, the Doppler period of Zak-OTFS modulation is constrained to be an integer multiple of the CP-OFDM sub-carrier spacing (SCS). Despite these constraints, simulations for the standardized non-line-of-sight (NLOS) Tapped Delay Line (TDL)-C channel (see \cite{TR38901}) reveal that at high Doppler spreads, the effective spectral efficiency (SE) performance of the proposed Zak-OTFS over CP-OFDM is significantly better than that achieved with CP-OFDM. However the effective SE achieved with Zak-OTFS over CP-OFDM is inferior to that achieved with unconstrained Zak-OTFS (where we are free to choose the delay and Doppler period and the pulse shaping and matched filters).

For example, numerical simulations in Section \ref{numsec} for the TDL-C channel (Urban-macro, normal delay profile and a maximum Doppler shift of $1.25$ KHz) for a packet having $1$ ms duration and bandwidth $720$ KHz, at a total received signal power to noise power ratio (TSNR) of $14$ dB (including data and pilots), reveal that the effective SE of Zak-OTFS over CP-OFDM (after considering the pilot and CP overheads) is $40$ percent more than that achieved with CP-OFDM.
Therefore, even with existing CP-OFDM based modems it is possible to achieve a significant improvement in the effective SE by simply implementing Zak-OTFS over CP-OFDM using the proposed architecture. 
The effective SE of unconstrained Zak-OTFS is about $6 \%$ higher than that of Zak-OTFS over CP-OFDM. For higher channel Doppler spreads, the gains are even higher. As an example, for a maximum Doppler shift of $6.48$ KHz and a line-of-sight (LOS) TDL-D channel (Urban-macro, short delay profile \cite{TR38901}), the effective SE of CP-OFDM, Zak-OTFS over CP-OFDM and unconstrained Zak-OTFS at a TSNR of $14$ dB are respectively, $0.56$, $1.33$ and $1.92$ bits/sec/Hz. Note that the effective SE achieved with unconstrained Zak-OTFS is roughly $30$ percent more than that achieved with Zak-OTFS over CP-OFDM.
A maximum Doppler shift of $1.25$ KHz is possible for a carrier frequency of $15$ GHz and a mobile speed of $90$ Km/hr (vehicular speed). A higher speed of $1000$ Km/hr (e.g., aircraft-to-ground communication) at a carrier frequency of $7$ GHz results in a maximum Doppler shift of $6.48$ KHz.

We have also simulated a no-mobility scenario (zero Doppler shift) with a long delay profile (maximum delay spread of roughly $10 \mu s$). For the proposed Zak-OTFS over CP-OFDM, the highest SE is achieved when $M=1$ (i.e., minimum delay period) which is CP-OFDM itself. However, unconstrained Zak-OTFS achieves significantly better performance than CP-OFDM. This is because, inter-symbol interference (ISI) degrades the effective SE achieved by CP-OFDM as the CP duration (with 3GPP 5G NR numerology) is limited to $4.7 \mu s$ which is not sufficient for large delay profile scenarios. For example, at a total received signal to noise ratio of $14$ dB, Zak-OTFS (unconstrained) achieves almost double the effective SE achieved with CP-OFDM. This scenario highlights the importance of Zak-OTFS not just for high mobility scenarios but also for long delay scenarios. 

We also show that CP-OFDM is the same as the proposed Zak-OTFS over CP-OFDM with delay period equal to its minimum possible value, i.e., minimum delay period Zak-OTFS over CP-OFDM.
In other words, in the proposed Zak-OTFS over CP-OFDM architecture, we can switch dynamically between Zak-OTFS over CP-OFDM and CP-OFDM simply by changing the delay period.
The proposed Zak-OTFS over CP-OFDM is in fact a family of modulations (corresponding to different values of the delay period) with CP-OFDM as a special case corresponding to the minimum possible value of the delay period. Numerical simulations in Section \ref{numsec} reveal that for the considered channel scenarios, the proposed Zak-OTFS over CP-OFDM achieves highest effective SE when its delay period is maximum, i.e., minimum Doppler period Zak-OTFS over CP-OFDM. The proposed Zak-OTFS over CP-OFDM architecture can also be extended to filtered OFDM \cite{FilterOFDM} by choosing the time-window appropriately.

This paper is organized as follows. In Section \ref{secprelim}, we discuss CP-OFDM and Zak-OTFS modulation and de-modulation.
In Section \ref{cpofdmsubsec} we present the DFT and IDFT based implementation of CP-OFDM modulator and de-modulator respectively. In Section \ref{zakotfssubsec}, we present the traditional continuous DD domain processing based Zak-OTFS transceiver architecture.
We also present the equivalent and more practical DZT based Zak-OTFS architecture. The transceiver architectures presented in Section \ref{secprelim} are known,
but we still include proofs of the results (in appendix) for the sake of completeness. In Section \ref{zakoverofdmtx} we propose the Zak-OTFS over CP-OFDM transmitter architecture and the corresponding receiver architecture is proposed in Section \ref{zakoverofdmrx}.
In Section \ref{subseciorel} we discuss the I/O relation of the proposed Zak-OTFS over CP-OFDM and channel acquisition using DD domain pilots.
In Section \ref{subsecsalient}, we summarize the salient features and advantages of the proposed Zak-OTFS over CP-OFDM architecture. In Section \ref{numsec}
we present simulation results on the effective SE comparison between unconstrained Zak-OTFS, Zak-OTFS over CP-OFDM and CP-OFDM for the TDL-C and TDL-D channel models. 

\section{Preliminaries: Doubly-spread channel, CP-OFDM and Zak-OTFS modulation} 	
\label{secprelim}
In this section we briefly describe the channel model and the transceiver signal processing in
CP-OFDM and Zak-OTFS. 
A doubly-spread channel is characterized by 
a channel response which is both time and frequency selective.
For a channel input $x(t)$, the channel response/output $r(t)$ is given by \cite{Bello}
\begin{eqnarray}
    r(t) & \hspace{-3mm} = & \hspace{-3mm} \iint h_{\text{phy}}(\tau, \nu) \, x(t - \tau) \, e^{j 2 \pi \nu (t - \tau)} \, d\tau \, d\nu \, + \, n(t)
\end{eqnarray}where $h_{\text{phy}}(\tau, \nu)$ is the channel delay-Doppler (DD) spreading function, $n(t)$ is AWGN with power spectral density (PSD) $N_0$ Watt/Hz. A commonly used model for $h(\tau, \nu)$ is the multi-path model, where
\begin{eqnarray}
    h(\tau, \nu) & = & \sum\limits_{i=1}^{P} h_i \, \delta(\tau - \tau_i) \, \delta(\nu - \nu_i),
 \end{eqnarray}i.e., there are $P$ paths, with the $i$-th path having complex channel gain $h_i$, and induces delay and Doppler shift of $\tau_i$ and $\nu_i$ respectively. For this multi-path model, the output is given by
 \begin{eqnarray}
     r(t) & = & \sum\limits_{i=1}^P h_i \, x(t - \tau_i) \, e^{j 2 \pi \nu_i (t - \tau_i)} \, + \, n(t).
 \end{eqnarray}
\begin{figure*}[h]
\vspace{-9mm}
\centering
\includegraphics[width=16.8cm, height=5.9cm]{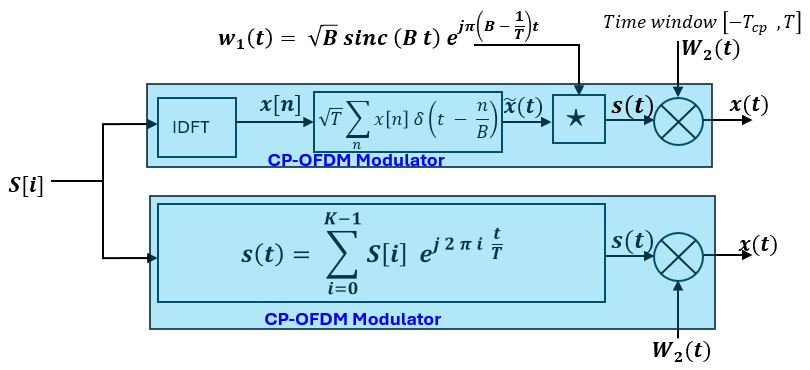}
\vspace{-2mm}
\caption{
Two equivalent implementations of CP-OFDM modulator based on, (i) inverse Discrete Fourier Transform (IDFT) (top chain), and (ii) superposition of continuous-time complex sinsuoidal carriers  (bottom chain).}
\label{fig0}
\vspace{-5mm}
\end{figure*}
\subsection{CP-OFDM modulator and demodulator}
\label{cpofdmsubsec}
Consider CP-OFDM with $K$ sub-carriers and carrier spacing $\Delta f$.
Let ${S}[i]$, $i=0,1,\cdots, K-1$ represent the $K$ frequency-domain (FD)
symbols, with the $i$-th FD symbol carried by a complex sinusoid of
frequency $i \Delta f$. The continuous-time (CT) information signal is therefore \cite{OFDMpaper, Cimini85}
\begin{eqnarray}
\label{steqn}
    s(t) & = &  \sum\limits_{i=0}^{K-1} {S}[i] \, e^{j 2 \pi i \Delta f t}.
\end{eqnarray}Note that $s(t)$ is periodic with period $T = 1/\Delta f$. Therefore, it suffices to transmit $s(t)$ restricted to a duration of $T$ seconds.
In CP-OFDM, we instead restrict $s(t)$ to a duration of $(T + T_{cp})$, where the cyclic repetition of $T_{cp}$ seconds acts as a guard time between two ``CP-OFDM symbols" and also simplifies the transceiver architecture for delay-only channels. Note that $s(t)$ is limited to the frequency-domain (FD) interval $[0 \,,\, (K-1) \Delta f]$. The transmitted CP-OFDM modulated signal is then given by
\begin{eqnarray}
    x(t) & = & s(t) \, W_2(t)
\end{eqnarray}where the time window $W_2(t)$ is given by
\begin{eqnarray}
\label{eqnW2t}
    W_2(t) & \Define  \begin{cases}
       \frac{1}{\sqrt{T}} &, -T_{cp} \leq t < T, \\
       0  &, \mbox{\small{otherwise}} \\
    \end{cases}.
\end{eqnarray}Due to time-windowing, $x(t)$ is exactly time limited to the TD interval $[-T_{cp} \,,\, T]$ and is approximately band-limited to the frequency-domain (FD) interval $\left[ -\frac{\Delta f}{2} \,,\, K \Delta f -\frac{\Delta f}{2}\right]$ and therefore the signal bandwidth is approximately $B \Define K \Delta f$ Hz.

CP-OFDM modulator is illustrated through Fig.~\ref{fig0} where the bottom signal processing chain is what we have discussed above. We next show that the top chain in Fig.~\ref{fig0} is an equivalent Inverse Discrete Fourier Transform (IDFT) based implementation of CP-OFDM. In the IDFT based implementation, IDFT of $S[i], i=0,1,\cdots, K-1$ gives the discrete-time signal
\begin{eqnarray}
\label{xneqn19834}
    x[n] & = & \mbox{\small{IDFT}}(S[i]) \nonumber \\
    & = & \frac{1}{\sqrt{K}} \sum\limits_{i=0}^{K-1} S[i] \, e^{j 2 \pi n \frac{i}{K}}\,\,,\,\, n \in {\mathbb Z}.
\end{eqnarray}Note that $x[n]$ is periodic with period $K$.
$x[n]$ linearly modulates a Dirac-delta pulse train resulting in the continuous-time signal
\begin{eqnarray}
    {\Tilde x}(t) & = & \sqrt{T} \sum\limits_{n \in {\mathbb Z}} x[n] \, \delta\left(t - \frac{n}{B} \right).
\end{eqnarray}This is followed by filtering with $w_1(t)$
followed by time-windowing with $W_2(t)$. The following result shows the equivalence of the two chains in Fig.~\ref{fig0}.

\begin{result}
\label{res01}
    Consider $w_1(t)$ given by
    \begin{eqnarray}
\label{eqnw1t}
    w_1(t) & = & \sqrt{B} \, sinc(B t) \, e^{j \pi \left( B - \frac{1}{T}\right)t},
\end{eqnarray}
    such that its Fourier transform is limited to $\left[ - \frac{\Delta f}{2} \,,\, K \Delta f - \frac{\Delta f}{2} \right]$, i.e.
    \begin{eqnarray}
    \label{eqnW1f}
        W_1(f) & \Define & \int w_1(t) \, e^{-j 2 \pi f t} \, dt \nonumber \\
        & = & \begin{cases}
            \frac{1}{\sqrt{K \Delta f}} \,,\, - \frac{\Delta f}{2} \leq f < K \Delta f - \frac{\Delta f}{2} \\
            0 \,,\, \mbox{\small{otherwise}} \\
        \end{cases}.
    \end{eqnarray}
    The filtering of ${\Tilde x}(t)$ (in the top chain of Fig.~\ref{fig0}) with $w_1(t)$ is equal to $s(t)$ given by (\ref{steqn}), i.e.
    \begin{eqnarray}
        {\Tilde x}(t) \, \star \, w_1(t) & = & s(t).
    \end{eqnarray}

\end{result}
\begin{IEEEproof}
See Appendix \ref{prfres01}.
\end{IEEEproof}

\begin{figure*}[h]
\vspace{-0mm}
\centering
\includegraphics[width=13.8cm, height=3.9cm]{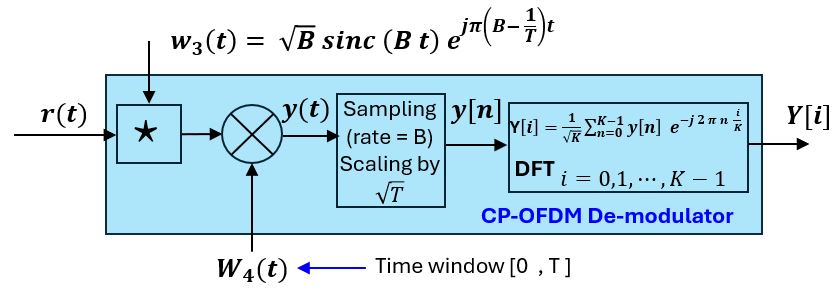}
\vspace{-2mm}
\caption{
A typical CP-OFDM demodulator implementation.}
\label{fig01}
\vspace{-5mm}
\end{figure*}

With i.i.d. zero mean FD symbols $S[i]$ each having variance ${\mathbb E}[ \vert S[i] \vert^2] = E$, the average power of the transmitted signal $x(t)$ is $E$.

A typical CP-OFDM demodulator implementation is shown in Fig.~\ref{fig01}.
At the receiver, we first pass the received signal $r(t)$ through a low-pass filter to restrict it to the FD interval $\left[ -\frac{\Delta f}{2} \,,\, B -\frac{\Delta f}{2}\right]$. Since $w_3(t)$ is normalized (total energy is unity), the AWGN power at the filter output is $N_0$. With normalized channel gains $h_i, i=1,2,\cdots, P$ (i.e., $\sum\limits_{i=1}^P {\mathbb E}[ \vert h_i \vert^2] = 1$), the average received signal power is equal
to the average transmit power $E$. Therefore, the signal to noise ratio (SNR) at the receiver is $E/N_0$. The filtered received signal is
\begin{eqnarray}
r(t) \, \star \, w_3(t),
\end{eqnarray}where $\star$ denotes linear convolution and the Fourier transform of $w_3(t)$ is given by
\begin{eqnarray}
    W_3(f) & = & \int w_3(t) \, e^{-j 2 \pi f t} \, dt \nonumber  \\
    & = & \begin{cases}
     \frac{1}{\sqrt{K \Delta f}} , \,\,\, -\frac{\Delta f}{2} \leq f < B - \frac{\Delta f}{2}, \\
     0 , \,\,\, \mbox{\small{otherwise}}, \\
    \end{cases},
    \end{eqnarray}which implies that
    \begin{eqnarray}
        \label{w3teqn92647}
        w_3(t) & = & \sqrt{B} \, sinc( B t) \, e^{j \pi \left( B - \frac{1}{T}\right) t}.
    \end{eqnarray}The filtered signal is then restricted in time to the TD interval $[0 \,,\, T]$ (i.e., removal of CP) resulting in the signal
    \begin{eqnarray}
    \label{w4teqn}
        y(t) & = & W_4(t) \, \left[ w_3(t) \, \star \, r(t) \right], \nonumber \\
        W_4(t) & = & \begin{cases}
     \frac{1}{\sqrt{T}} , \,\,\, 0 \leq t < T, \\
     0 \,, \,\,\, \mbox{\small{otherwise}}, \\
    \end{cases}.
    \end{eqnarray}
    $y(t)$ is then sampled at a sampling rate of $B$ Hz and scaled by $\sqrt{T}$ to give the discrete-time signal
    \begin{eqnarray}
        y[n] & = & \sqrt{T} \, y\left( t = \frac{n}{B}\right).
    \end{eqnarray}The received FD symbols on the $K$ sub-carriers are then given by the Discrete Fourier Transform (DFT) of the $K$ discrete-time
    samples, $y[0], y[1], \cdots, y[K-1]$, i.e.
    \begin{eqnarray}
        {Y}[i] & = & \frac{1}{\sqrt{K}} \sum\limits_{n=0}^{K-1} y[n] \, e^{-j 2 \pi n \frac{i}{K}},
    \end{eqnarray}$i=0,1,2,\cdots, K-1$.
    In the absence of channel Doppler spread, ${Y}[i]$ depends only on $S[i]$ and therefore per-carrier FD equalization suffices. However, in the presence of channel Doppler spread, ${ Y}[i]$ consists also of interference from symbols transmitted on other carriers i.e., $S[q], q \ne i$
    which degrades the performance of per-carrier equalization.
    
\subsection{Zak-OTFS modulator and demodulator}
\label{zakotfssubsec}
We present Zak-OTFS modulation as discussed in \cite{ZAKOTFS1, ZAKOTFS2, otfsbook}.
Let us consider Zak-OTFS modulation with information lattice
\begin{eqnarray}
\label{lambdap}
    \Lambda_p & \Define & \left\{ \left(\frac{k}{B}, \frac{l}{T} \right) \, \vert \, k, l \in {\mathbb Z} \right\}.
\end{eqnarray}Let $(\tau_p, \nu_p)$ ($\tau_p \, \nu_p = 1$) be the Zak-OTFS period parameters, such that $M \Define B \tau_p$ and $N \Define T \nu_p$ are integers.
The $MN= BT$ information symbols $x[k,l]$, $k=0,1,\cdots, M-1$, $l=0,1,\cdots, N-1$
are embedded into the discrete DD domain quasi-periodic signal $x_{dd}[k,l]$ as
\begin{eqnarray}
    x_{dd}[k,l] & = & x[k \, \text{mod} \, M,l \, \text{mod} \, N] \, e^{j 2 \pi \lfloor \frac{k}{M} \rfloor \frac{l}{N}},
\end{eqnarray}$k,l \in {\mathbb Z}$. Note that $x_{dd}[k,l]$ is quasi-periodic with periods
$M$ and $N$ respectively along the discrete delay and Doppler axes, i.e. for all $k,l,n,m \in {\mathbb Z}$
\begin{eqnarray}
\label{qpeqn234}
    x_{dd}[k + nM, l + mN] & = & e^{j 2 \pi n \frac{l}{N}} \, x_{dd}[k,l].
\end{eqnarray}
This is then embedded into a continuous DD signal

{\vspace{-4mm}
\begin{eqnarray}
\label{eqn3494}
    x_{dd}(\tau, \nu) & = & \sum\limits_{k,l \in {\mathbb Z}} x_{dd}[k,l] \delta(\tau - k/B) \, \delta(\nu - l/T),
\end{eqnarray}\normalsize}which is then filtered with the pulse shaping filter
$w_{tx}(\tau, \nu)$ resulting in the filtered DD signal
\begin{eqnarray}
\label{pulseshapingtx}
    x_{dd}^{w_{tx}}(\tau, \nu) & = & w_{tx}(\tau, \nu) \, *_{\sigma} \, x_{dd}(\tau, \nu),
\end{eqnarray}where $*_{\sigma}$ denotes the twisted convolution operation.
The TD realization of the filtered DD signal is the Zak-OTFS modulated transmit signal
\begin{eqnarray}
\label{eqnxtzak}
    x(t) & = & {\mathcal Z}_t^{-1}\left(  x_{dd}^{w_{tx}}(\tau, \nu) \right) \nonumber \\
    & = & \sqrt{\tau_p} \int\limits_{0}^{\nu_p} x_{dd}^{w_{tx}}(t, \nu) \, d\nu,
\end{eqnarray}where ${\mathcal Z}_t^{-1}$ denotes the inverse Zak transform \cite{otfsbook}. At the receiver, the DD domain realization of the received signal $r(t)$
is given by its Zak-transform \cite{otfsbook}
\begin{eqnarray}
    r_{dd}(\tau, \nu) & = & {\mathcal Z}_t\left( r(t) \right) \nonumber \\
    & = & \sqrt{\tau_p} \sum\limits_{n \in {\mathbb Z}} r(\tau + n \tau_p) \, e^{-j 2 \pi n \nu \tau_p}.
\end{eqnarray}This is then match-filtered with the DD filter $w_{rx}(\tau, \nu)$
resulting in the filtered DD domain received signal
\begin{eqnarray}
\label{yddeqn183}
    y_{dd}(\tau, \nu) & = & w_{rx}(\tau, \nu) *_{\sigma} r_{dd}(\tau, \nu).
\end{eqnarray}This is then sampled on the information lattice $\Lambda_p$ resulting in the received discrete DD domain signal
\begin{eqnarray}
\label{eqnyddkl}
    y_{dd}[k,l] & = & y_{dd}\left( \tau = \frac{k}{B} \,,\, \nu = \frac{l}{T}\right).
\end{eqnarray}This discrete DD domain signal is quasi-periodic with periods $M$ and $N$ along the discrete delay and Doppler axis respectively, i.e., for all $k,l,n,m \in {\mathbb Z}$
\begin{eqnarray}
y_{dd}[k+nM,l+mN] & = & e^{j 2 \pi n \frac{l}{N}} \, y_{dd}[k,l].
\end{eqnarray}Since the DD carriers interfere in a doubly-spread channel,
joint detection of all $MN$ information symbols is carried out from the received
discrete DD domain signal in one period, i.e., from $y_{dd}[k,l]$, $k=0,1,\cdots, M-1$, $l=0,1,\cdots, N-1$. Recently, joint equalization/detection techniques have been proposed for Zak-OTFS having low complexity ${\mathcal O}(M^2 N^2)$ only \cite{lowcomplexityzakotfseq}.

The Zak-OTFS I/O relation is then given by \cite{ZAKOTFS2, otfsbook}
\begin{eqnarray}
\label{eqniorel1}
    y_{dd}[k,l] & = & h_{dd}[k,l] \, *_{\sigma} \, x_{dd}[k,l] \, + \, n_{dd}[k,l],
\end{eqnarray}where $n_{dd}[k,l]$ is the Zak transform of AWGN $n(t)$ match-filtered with $w_{rx}(\tau, \nu)$ followed by sampling on the information lattice $\Lambda_p$, $*_{\sigma}$ denotes the twisted convolution operation given by
\begin{eqnarray}
\label{eqniorel2}
    h_{dd}[k,l] \, *_{\sigma} \, x_{dd}[k,l] &  & \nonumber \\
    & & \hspace{-25mm} = \sum\limits_{k',l' \in {\mathbb Z}} h_{dd}[k',l'] \, x_{dd}[k - k', l - l'] \, e^{j 2 \pi l' \frac{(k - k')}{MN}}.
\end{eqnarray}Also, $h[k,l]$ is the discrete DD domain effective channel filter given by
\begin{eqnarray}
\label{eqniorel3}
    h_{dd}[k,l] & \Define & h\left( \tau = \frac{k}{B} \,,\, \nu = \frac{l}{T}\right), \nonumber \\
    h_{dd}(\tau, \nu) & \Define & w_{rx}(\tau, \nu) \, *_{\sigma} \, h_{\text{phy}}(\tau, \nu) \, *_{\sigma} \, w_{tx}(\tau, \nu).
\end{eqnarray}Therefore, knowledge of $h_{dd}[k,l]$ is sufficient to predict the channel response to any channel input. It has been shown in \cite{ZAKOTFS2, otfsbook} that
$h_{dd}[k,l]$ can be acquired efficiently by using a single pilot carrier in the DD domain if the crystallization condition is satisfied, i.e., the channel delay spread is less than the delay period $\tau_p$ and the channel Doppler spread is less than the Doppler period $\nu_p$. Therefore, for doubly-spread channels the Zak-OTFS I/O relation is predictable when the crystallization condition is satisfied, since the channel response to a particular DD carrier (e.g., a single pilot DD carrier) can be used to accurately predict the channel response to any other arbitrary DD carrier. {This
is however not the case with \emph{CP-OFDM which is not predictable in doubly-spread channels}.}
\begin{figure*}[h]
\vspace{-9mm}
\centering
\includegraphics[width=16.8cm, height=5.9cm]{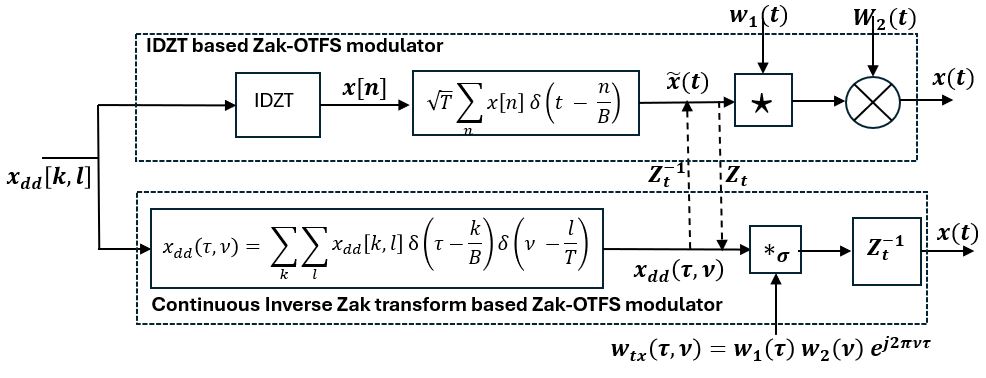}
\vspace{-2mm}
\caption{
Two equivalent implementations of Zak-OTFS modulator, (i) inverse Discrete Zak Transform (IDZT) based processing (top chain), and (ii) continuous inverse Zak transform (${\mathcal Z}_t^{-1}$) based processing (bottom chain).}
\label{fig1}
\vspace{-1mm}
\end{figure*}

\subsection{Discrete Zak Transform (DZT) based implementation of Zak-OTFS}
In this section, we present a DZT based implementation of
Zak-OTFS and show that it is equivalent to the continuous Zak transform
based implementation presented in Section \ref{zakotfssubsec}. 

As shown in Fig.~\ref{fig1}, there are two signal processing chains,
both taking as input $x_{dd}[k,l]$ and generating the continuous-time
transmit signal $x(t)$.
The bottom chain in the figure is based on the continuous
DD domain inverse Zak transform. We now describe the top chain which is based
on the Inverse Discrete Zak Transform (IDZT) and is a more practical implementation of Zak-OTFS modulation (see Appendix A in Chapter $8$ of \cite{otfsbook}).

In IDZT based Zak-OTFS modulation, $x_{dd}[k,l]$ is converted to a discrete-time (DT) periodic signal $x[n]$ using IDZT, i.e.,
\begin{eqnarray}
\label{idzteqn92367}
    x[n] & = & \frac{1}{\sqrt{N}} \sum\limits_{l=0}^{N-1} x_{dd}[n,l], \,\, n \in {\mathbb Z}.
\end{eqnarray}Note that $x[n] = x[n + MN]$ for all $n \in {\mathbb Z}$, i.e.,
$x[n]$ is periodic with period $MN$, since $x_{dd}[k,l]$ is quasi-periodic (see (\ref{qpeqn234})). Next, $x[n]$ is converted to the continuous-time (CT) signal
\begin{eqnarray}
    {\Tilde x}(t) & \Define & \sqrt{T} \sum\limits_{n \in {\mathbb Z}} x[n] \, \delta\left(t - \frac{n}{B} \right).
\end{eqnarray}This is then followed by pulse shaping, i.e., linear convolution with $w_1(t)$ to limit the bandwidth of ${\Tilde x}(t)$ followed by time-windowing with $W_2(t)$ to limit the time duration of ${\Tilde x}(t)$. The transmitted signal is
\begin{eqnarray}
\label{idztxeqn}
  W_2(t) \, \left[  w_1(t) \, \star \, {\Tilde x}(t)\right].
\end{eqnarray}The following result shows that this transmit signal is same as $x(t)$ in (\ref{eqnxtzak}) with a particular choice of the pulse shaping filter $w_{tx}(\tau, \nu)$.
\begin{figure*}[h]
\vspace{-1mm}
\centering
\includegraphics[width=16.8cm, height=5.9cm]{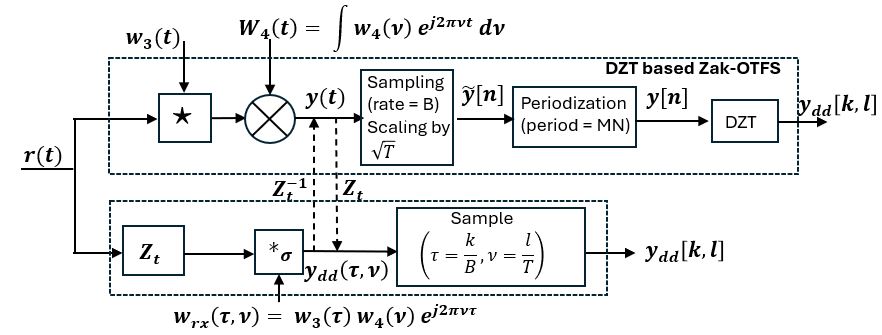}
\vspace{-2mm}
\caption{
Two equivalent implementations of Zak-OTFS demodulator, (i) Discrete Zak Transform (DZT) based processing (top chain), and (ii) continuous Zak transform (${\mathcal Z}_t^{-1}$) based processing (bottom chain).}
\label{fig2}
\vspace{-5mm}
\end{figure*}
\begin{result}[see Chapter $8$ (Appendix A) in \cite{otfsbook} and also \cite{lowcomplexityzakotfseq}]
\label{res1}
The two signal processing chains for Zak-OTFS modulation in Fig.~\ref{fig1} are equivalent, i.e., the transmit signal of the continuous inverse Zak transform
based processing is same as that of IDZT based processing
\begin{eqnarray}
    x(t) & = &  W_2(t) \, \left[  w_1(t) \, \star \, {\Tilde x}(t)\right],
\end{eqnarray}when the DD domain transmit pulse shaping filter in the bottom chain is chosen to be
\begin{eqnarray}
\label{wtxeqn1937}
    w_{tx}(\tau, \nu) & = & w_1(\tau) \, w_2(\nu) \, e^{j 2 \pi \nu \tau}, \nonumber \\
    w_2(\nu) & \Define & \int W_2(t) \, e^{-j 2 \pi \nu t} \, dt.
\end{eqnarray}
\end{result}
\begin{IEEEproof}
See Appendix \ref{prfres1}.
\end{IEEEproof}

Next, we show that the continuous Zak transform based Zak-OTFS demodulator discussed in Section \ref{zakotfssubsec} is equivalent to DZT based demodulator shown in the top signal processing chain in Fig.~\ref{fig2}.
In the DZT based Zak-OTFS demodulator in Fig.~\ref{fig2}, the received TD signal $r(t)$ is filtered with $w_3(t)$ followed by time-windowing with $W_4(t)$ resulting in the signal
\begin{eqnarray}
\label{yteqn2864}
    y(t) & = &  W_4(t) \, \left[  w_3(t) \, \star \, r(t) \right]
\end{eqnarray}which is then sampled at a sampling rate of $B$ Hz resulting in the discrete-time signal ${\Tilde y}[n]$, i.e.
\begin{eqnarray}
\label{tildeyneqn}
{\Tilde y}[n] & = & {\sqrt{T}} \, y\left( t = \frac{n}{B} \right).
\end{eqnarray}This is then periodized with period $MN$ resulting in the
$MN$-periodic signal
\begin{eqnarray}
\label{periodizeeqn}
    y[n] & = & \sum\limits_{p \in {\mathbb Z}} {\Tilde y}[n + pMN].
\end{eqnarray}This is followed by DZT of $y[n]$, which the next result shows is equal to the discrete DD domain signal $y_{dd}[k,l]$ at the output of the continuous Zak transform based Zak-OTFS demodulator (see (\ref{eqnyddkl})).

\begin{result}
\label{res31}
    The output $y_{dd}[k,l]$ of the continuous Zak transform based demodulator (bottom chain of Fig.~\ref{fig2}) with the DD
    domain matched filter
    \begin{eqnarray}
    \label{eqn826547}
        w_{rx}(\tau, \nu) & = & w_3(\tau) \, w_4(\nu) \, e^{j 2 \pi \nu \tau}
    \end{eqnarray}is equal to the DZT of the sampled and periodized discrete-time signal in the DZT based demodulator (top chain in Fig.~\ref{fig2}), i.e.,
    \begin{eqnarray}
    \label{dzteqn2974}
      y_{dd}[k,l] & = & \frac{1}{\sqrt{N}} \sum\limits_{q=0}^{N-1} y[k + qM] \, e^{-j 2 \pi \frac{q l}{N}},
    \end{eqnarray}with the time-windowing waveform $W_4(t)$ equal to the TD realization of $w_4(\nu)$, i.e.
    \begin{eqnarray}
    \label{w4teqn1}
        W_4(t) & = & \int w_4(\nu) \, e^{j 2 \pi \nu t} \, d\nu.
    \end{eqnarray}
\end{result}
\begin{IEEEproof}
See Appendix \ref{prfres31}.
\end{IEEEproof}

\begin{figure}[h]
\vspace{-2mm}
\centering
\includegraphics[width=8.8cm, height=2.9cm]{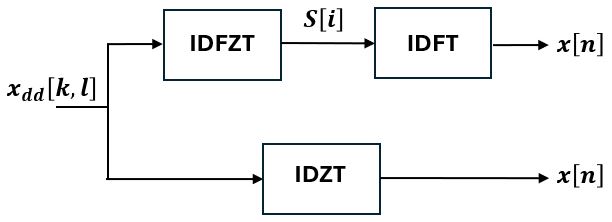}
\vspace{-2mm}
\caption{
IDZT as composition of IDFZT followed by IDFT.}
\label{fig3}
\vspace{-5mm}
\end{figure}

\begin{figure*}[h]
\vspace{-9mm}
\centering
\includegraphics[width=16.8cm, height=6.9cm]{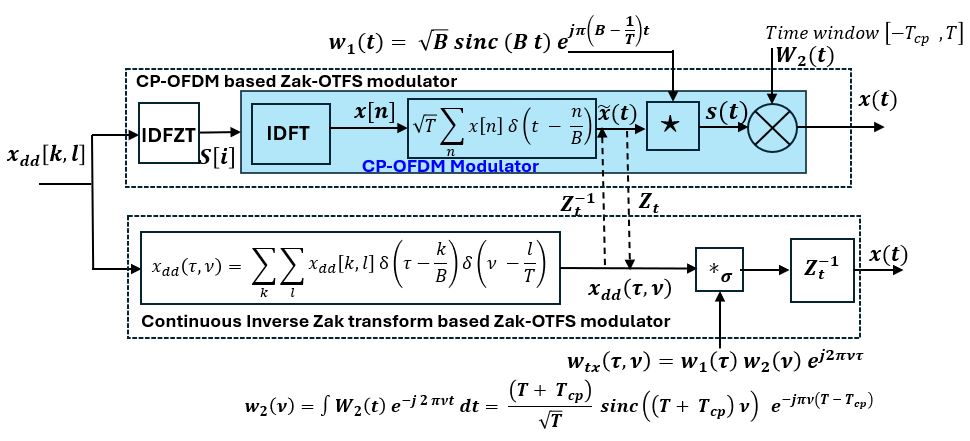}
\vspace{-2mm}
\caption{
Zak-OTFS modulation implemented as IDFZT followed by CP-OFDM modulator (see top chain).}
\label{fig5}
\vspace{-5mm}
\end{figure*}
\section{Zak-OTFS modulation as a precoder over CP-OFDM}
\label{zakoverofdmtx}
We note the striking resemblance between the IDZT based implementation of Zak-OTFS modulation in Fig.~\ref{fig1} (top chain) and the IDFT based implementation of CP-OFDM in Fig.~\ref{fig0} (top chain). The only difference is that we have IDZT in Fig.~\ref{fig1} and IDFT in Fig.~\ref{fig0}, and also in case of CP-OFDM the low-pass filter $w_1(t)$ and the time-window $W_2(t)$ are constrained. The following result shows that IDZT can be expressed as a composition of Inverse Discrete Frequency Zak Transform (IDFZT) followed by IDFT (see Fig.~\ref{fig3}). Replacing IDZT in the top chain of Fig.~\ref{fig1} with IDFZT followed by IDFT,
the signal processing from the IDFT input to the output signal $x(t)$ is exactly same as that of the IDFT based CP-OFDM modulator in the top chain of Fig.~\ref{fig0}. This implies that Zak-OTFS (with restrictions on $w_1(t)$ and $W_2(t)$) can be implemented as a \emph{precoder} over CP-OFDM, i.e., the DD domain signal $x_{dd}[k,l]$ is converted to its discrete FD representation $S[i]$ using IDFZT. $S[i]$ is then input to the CP-OFDM demodulator which generates the modulated signal $x(t)$.

\begin{result}
\label{res4}
    As shown in Fig.~\ref{fig3}, consider a discrete DD domain signal $x_{dd}[k,l]$ having delay and Doppler period $M$ and $N$ respectively. IDZT of $x_{dd}[k,l]$ gives its discrete-time representation $x[n]$, i.e.
    \begin{eqnarray}
    \label{idzteqn9364}
        x[n] & = & \text{IDZT}(x_{dd}[k,l]) \nonumber \\
        &  =  & \frac{1}{\sqrt{N}} \sum\limits_{l=0}^{N-1} x_{dd}[n,l], \,\,\, n \in {\mathbb Z}.
    \end{eqnarray}This conversion from $x_{dd}[k,l]$ to $x[n]$ can also be achieved by first converting $x_{dd}[k,l]$ to FD symbols $S[i]$ using IDFZT (see Appendix of Chapter $8$ in \cite{otfsbook}), i.e.
    \begin{eqnarray}
\label{sieqn12}
    S[i] & = & \text{IDFZT}(x_{dd}[k,l]) \nonumber \\
    & = & \frac{1}{\sqrt{M}} \sum\limits_{k=0}^{M-1} x_{dd}[k,i] \, e^{-j 2 \pi \frac{i k}{MN}}, \,\,i \in {\mathbb Z},
\end{eqnarray}followed by converting $S[i]$ to $x[n]$ using IDFT, i.e.
    \begin{eqnarray}
    \label{idft92784}
        x[n] & = & \text{IDFT}(S[i]) \nonumber \\
        & = & \frac{1}{\sqrt{MN}} \, \sum\limits_{i=0}^{MN-1} S[i] \, e^{j 2 \pi n \frac{i}{MN}}.
    \end{eqnarray}
\end{result}
\begin{IEEEproof}
See Appendix \ref{prfres4}.
\end{IEEEproof}

\begin{theorem}
\label{thm1}
    The top chain in Fig.~\ref{fig5} implements Zak-OTFS as a precoder over CP-OFDM,
    with Zak-OTFS parameters $M$ and $N$ being any positive integers
    such that $M N = K$ (no. of CP-OFDM sub-carriers), and the DD domain transmit pulse shaping filter is given by
    \begin{eqnarray}
    \label{eqnwtxzakoverofdm}
        w_{tx}(\tau, \nu) & = & w_1(\tau) \, w_2(\nu) \, e^{j 2 \pi \nu \tau}, \nonumber \\
        w_1(\tau) & = & \sqrt{B} \, sinc(B \tau) \, e^{j \pi \left(B - \frac{1}{T} \right) \tau}, \nonumber \\
        w_2(\nu) & = & \frac{(T + T_{cp})}{\sqrt{T}} \, sinc((T + T_{cp}) \nu) \, e^{-j \pi \nu \left( T - T_{cp}\right) }. \nonumber \\
    \end{eqnarray}Note that the corresponding delay and Doppler period of Zak-OTFS over CP-OFDM is given by
    \begin{eqnarray}
     \nu_p & = & \frac{N}{T} \, = \, N \Delta f  \,,\, \tau_p = \frac{1}{\nu_p} \, = \, \frac{M}{B}.    
    \end{eqnarray}The Zak-OTFS precoding operation is simply IDFZT in (\ref{sieqn12}), which takes as input $x_{dd}[k,l]$
    and gives the FD symbols $S[i]$ as output.
\end{theorem}
\begin{IEEEproof}
IDZT can be expressed as a composition of Inverse Discrete Frequency Zak Transform (IDFZT) followed by IDFT, as shown in Fig.~\ref{fig3}.
By substituting the IDZT module in the top chain of Fig.~\ref{fig1} with an IDFZT followed by IDFT, we get the Zak-OTFS modulator in top chain in Fig.~\ref{fig5},
where the top and the bottom chain are equivalent Zak-OTFS modulators.
In the top chain in Fig.~\ref{fig5}, it is clear that the signal processing after IDFZT (i.e., from the input of IDFT to the output $x(t)$) is exactly same as that of the CP-OFDM modulator in the top chain of Fig.~\ref{fig0}. However, since the filter $w_1(t)$ and the time-window $W_2(t)$ in CP-OFDM are constrained to the sinc pulse and the rectangular window respectively (see (\ref{eqnW2t}), (\ref{eqnw1t}) and Fig.~\ref{fig0}), the Zak-OTFS transmit pulse shaping filter in the equivalent bottom chain of Fig.~\ref{fig5} is therefore constrained to $w_{tx}(\tau, \nu)$ in (\ref{eqnwtxzakoverofdm}).

Also, from the composition of IDZT into IDFZT and IDFT (see Result \ref{res4}), it follows that the number of CP-OFDM carriers $K$ should be equal to the size of IDFT which is $MN$. Clearly, the time-duration of the Zak-OTFS packet/frame will be same as that of the CP-OFDM symbol which is $(T + T_{cp})$ and the bandwidth will be equal to that of CP-OFDM i.e., $B = K \Delta f = MN \Delta f$ Hz. 
\end{IEEEproof}

\textbf{Complexity of Precoding Zak-OTFS over CP-OFDM:}
As shown in Fig.~\ref{fig5}, additional complexity of precoding Zak-OTFS over CP-OFDM requires computing the IDFZT of $x_{dd}[k,l]$. From (\ref{sieqn12})
it is clear that $S[i]$ is periodic with period $MN$ since
\begin{eqnarray}
    S[i + MN] & \hspace{-3mm} = & \hspace{-3mm} \frac{1}{\sqrt{M}} \sum\limits_{k=0}^{M-1} \hspace{-2mm} x_{dd}[k, i+MN] \, e^{-j 2 \pi \frac{(i+MN)k}{MN}} \nonumber \\
    & \hspace{-3mm} = & \hspace{-3mm} \frac{1}{\sqrt{M}} \sum\limits_{k=0}^{M-1} \hspace{-2mm} x_{dd}[k, i] \, e^{-j 2 \pi \frac{ik}{MN}},
\end{eqnarray}where the last step follows from the periodicity of $x_{dd}[k,l]$ along the Doppler axis with period $N$. Therefore, it suffices to compute
$S[i]$, only for $i=0,1,\cdots, MN-1$.

In the following we describe how $S[i], i=0,1,\cdots, MN-1$
can be computed from $x_{dd}[k,l]$, $k =0,1,\cdots, M-1$, $l=0,1,\cdots, N-1$.
Since $x_{dd}[k,l]$ is periodic along the Doppler axis with period $N$, for all
$p \in {\mathbb Z}$, we have

{\vspace{-4mm}
\small
\begin{eqnarray}
\label{eqnddspread}
S[i+pN] 
&  \hspace{-3mm} =  & \hspace{-3mm}   \frac{1}{\sqrt{M}} \sum\limits_{k=0}^{M-1} x_{dd}[k, i]  \, e^{-j 2 \pi \frac{(i + pN) k}{MN}} ,
\end{eqnarray}\normalsize}i.e., for each $i=0,1,\cdots, N-1$, the $M$ DD symbols $x_{dd}[k,i], k=0,1,\cdots, M-1$ are jointly precoded across the $M$ OFDM subcarriers $S[(i + pN) \, \mbox{\small{mod}}  \, MN], p=0,1,\cdots, M-1$. For any given $i \in \{0,1,\cdots, N-1 \}$, let $z_i[k] \Define x_{dd}[k,i] e^{-j 2 \pi i k /(MN)}, \,\, k , i \in {\mathbb Z}$, and let $ S_i[p]  \Define  S[i+ pN], p =0,1,\cdots, M-1$. Then from (\ref{eqnddspread}) we get
\begin{eqnarray}
   S_i[p] & \hspace{-3mm} = & \hspace{-3mm} \frac{1}{\sqrt{M}} \hspace{-1mm} \sum\limits_{k=0}^{M-1} \hspace{-1mm} z_i[k]  \, e^{-j 2 \pi \frac{p k}{M}} \,,\,\, p=0,1,\cdots, M-1,
\end{eqnarray}which can be computed with $O(M \log(M))$ complexity using $M$-point Fast Fourier Transform (FFT). Since $i=0,1,\cdots, N-1$, the complexity of computing $S[i], i=0,1,\cdots, MN-1$ is therefore ${\mathcal O}(MN\log(M))$.

\begin{figure*}[h]
\vspace{-9mm}
\centering
\includegraphics[width=16.8cm, height=6.9cm]{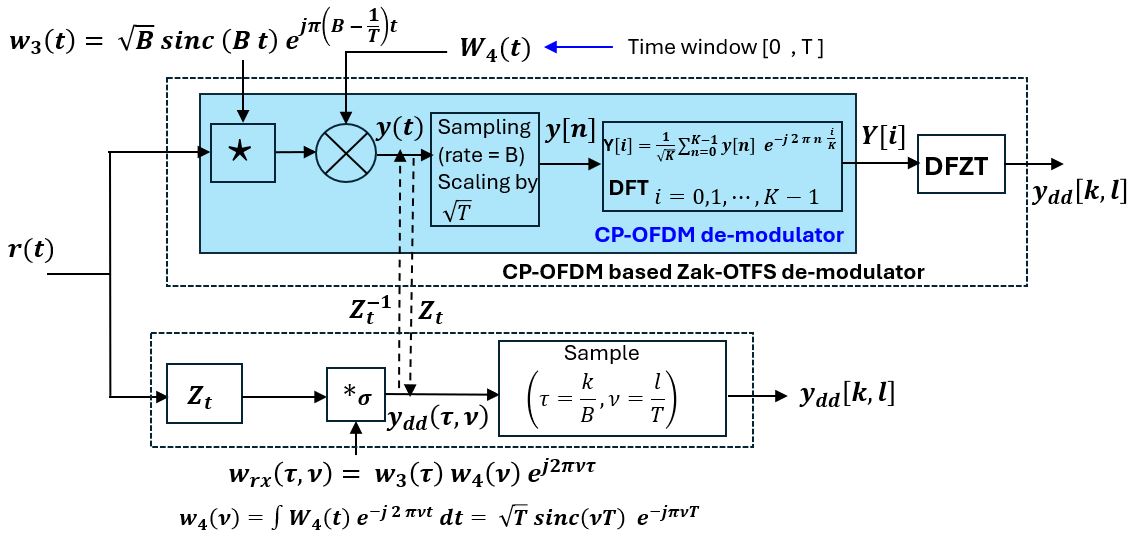}
\vspace{-2mm}
\caption{
Zak-OTFS de-modulation implemented as CP-OFDM de-modulator followed by DFZT (see top chain).}
\label{fig6}
\vspace{-5mm}
\end{figure*}
\section{Zak-OTFS de-modulation as post-processing after CP-OFDM de-modulation}
\label{zakoverofdmrx}

\begin{figure}[h]
\vspace{-2mm}
\centering
\includegraphics[width=8.8cm, height=2.9cm]{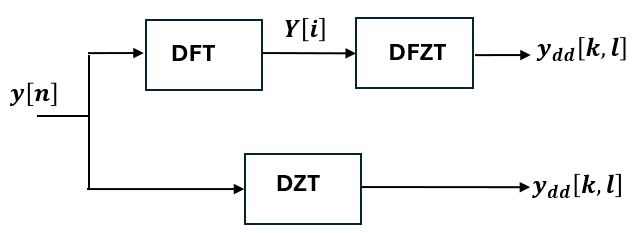}
\vspace{-2mm}
\caption{
DZT as composition of DFT followed by DFZT.}
\label{fig4}
\vspace{-5mm}
\end{figure}
Note the striking resemblance between the DZT based implementation of Zak-OTFS de-modulation in Fig.~\ref{fig2} (top chain) and the DFT based implementation of CP-OFDM de-modulation in Fig.~\ref{fig01}. The only difference is that we have DZT in Fig.~\ref{fig2} and DFT in Fig.~\ref{fig01}, and also in case of CP-OFDM the low-pass filter $w_3(t)$ and the time-window $W_4(t)$ are fixed to the sinc filter and the rectangular window in (\ref{w3teqn92647}) and (\ref{w4teqn}) respectively. Result \ref{res5} in the following shows that DZT can be expressed as a composition of DFT followed by the Discrete Frequency Zak Transform (DFZT) (see Fig.~\ref{fig4}). Therefore, substituting DZT
in the Zak-OTFS de-modulator in Fig.~\ref{fig2} with DFT followed by DFZT leads us to the top chain in Fig.~\ref{fig6} where the signal processing from the input $r(t)$
till the DFT output $Y[i]$ is exactly same as the CP-OFDM de-modulator in Fig.~\ref{fig01} (with $w_3(t)$ and $W_4(t)$ given by (\ref{w3teqn92647}) and (\ref{w4teqn}) respectively). In other words, Zak-OTFS de-modulation can be implemented as a post-processing of the CP-OFDM demodulator output $Y[i]$, i.e., the Zak-OTFS received discrete DD domain signal $y_{dd}[k,l]$ is given by applying DFZT (post-processing) to $Y[i]$ (see Theorem \ref{thm2} later).

\begin{result}
\label{res5}
    As shown in Fig.~\ref{fig4}, consider a discrete-time periodic signal $y[n]$ with period $MN$ whose DD domain representation $y_{dd}[k,l]$ is given by its DZT, i.e.
    \begin{eqnarray}
    \label{dzteqn9364}
        y_{dd}[k,l] & = & \text{DZT}(y[n]) \nonumber \\
        &  =  & \frac{1}{\sqrt{N}} \sum\limits_{q=0}^{N-1} y[k+qM] \, e^{-j 2 \pi \frac{q l}{N}}, \,\,\,
    \end{eqnarray}$k,l \in {\mathbb Z}$. This conversion from $y[n]$ to $y_{dd}[k,l]$ can also be achieved by first converting $y[n]$ to its FD realization $Y[i]$ using
    DFT, i.e.
    \begin{eqnarray}
    \label{dft92784}
        Y[i] & = & \text{DFT}(y[n]) \nonumber \\
        & = & \frac{1}{\sqrt{MN}} \, \sum\limits_{n=0}^{MN-1} y[n] \, e^{-j 2 \pi n \frac{i}{MN}}, \,\, i \in {\mathbb Z},
    \end{eqnarray}followed by conversion of $Y[i]$ to its DD domain
    realization using Discrete Frequency Zak Transform (DFZT) (see Appendix of Chapter $8$ in \cite{otfsbook}), i.e.
    \begin{eqnarray}
\label{seqn12988}
    y_{dd}[k,l] & = & \text{DFZT}(Y[i]) \nonumber \\
    & = & \frac{1}{\sqrt{M}} \sum\limits_{p=0}^{M-1} Y[l + pN] \, e^{j 2 \pi (l + pN) \frac{k}{MN}},
\end{eqnarray}$k,l \in {\mathbb Z}$.
\end{result}
\begin{IEEEproof}
See Appendix \ref{prfres5}.
\end{IEEEproof}

The next theorem shows that the Zak-OTFS de-modulator can be implemented as a post-processing of the output of CP-OFDM demodulator (see Fig.~\ref{fig6}).
\begin{theorem}
\label{thm2}
    The top chain in Fig.~\ref{fig6} implements Zak-OTFS de-modulation as a post-processing of the output of the CP-OFDM de-modulator,
    with Zak-OTFS parameters $M$ and $N$ being any positive integers
    such that $M N = K$ (no. of CP-OFDM sub-carriers), and the DD domain receiver matched filter given by
    \begin{eqnarray}
    \label{eqnwrxzakoverofdm}
        w_{rx}(\tau, \nu) & = & w_3(\tau) \, w_4(\nu) \, e^{j 2 \pi \nu \tau}, \nonumber \\
        w_3(\tau) & = & \sqrt{B} \, sinc(B \tau) \, e^{j \pi \left(B - \frac{1}{T} \right) \tau}, \nonumber \\
        w_4(\nu) & = & \sqrt{T} \, sinc(\nu T) \, e^{-j \pi \nu T }. \nonumber \\
    \end{eqnarray}Note that the corresponding delay and Doppler period of Zak-OTFS over CP-OFDM is given by
    \begin{eqnarray}
     \nu_p & = & \frac{N}{T} \, = \, N \Delta f  \,,\, \tau_p = \frac{1}{\nu_p} \, = \, \frac{B}{M}.    
    \end{eqnarray}The Zak-OTFS post-processing operation is simply
    DFZT (in (\ref{seqn12988})), which takes as input $Y[i]$ and generates its DD domain representation $y_{dd}[k,l]$.
\end{theorem}
\begin{IEEEproof}
From Result \ref{res5} we know that DZT can be expressed as a composition of DFT followed by DFZT, as shown in Fig.~\ref{fig4}.
By substituting the DZT module in the top chain of Fig.~\ref{fig2} with DFT followed by DFZT, and choosing filter $w_3(t)$ and time-window $W_4(t)$ given by (\ref{w3teqn92647}) and (\ref{w4teqn}) respectively, we get the Zak-OTFS de-modulator in the top chain in Fig.~\ref{fig6}. Note that, 
$W_4(t)$ is a rectangular window waveform limited to the TD interval $[0 \,,\, T]$, and therefore $y(t)$ in Fig.~\ref{fig2} is zero outside the TD interval $[0 \,,\, T]$. Hence, for this choice of $W_4(t)$, the periodization module in the DZT based Zak-OTFS modulator in Fig.~\ref{fig2} is not required since ${\Tilde y}[n] = y[n], n=0,1,\cdots, MN-1$. Also, choosing $w_3(t)$ and $W_4(t)$ given by 
(\ref{w3teqn92647}) and (\ref{w4teqn}) respectively, ensures that the signal processing in the top chain of Fig.~\ref{fig6} starting from the input $r(t)$
to the output $Y[i]$ of the DFT is exactly same as the CP-OFDM demodulator in the top chain of Fig.~\ref{fig01}. Also, from Result \ref{res31} we know that the top and the bottom chains in Fig.~\ref{fig6} are equivalent Zak-OTFS de-modulators, and therefore the equivalent DD domain matched-filter is clearly given by (\ref{eqnwrxzakoverofdm}).

From the de-composition of DZT into DFT followed by DFZT (see Result \ref{res5}), it follows that the number of CP-OFDM carriers $K$ should be equal to the size of DFT which is $MN$.
\end{IEEEproof}

\textbf{Complexity of DFZT post-processing in the CP-OFDM based Zak-OTFS de-modulator:}
The received DD domain signal $y_{dd}[k,l]$ is given by DZFT of the CP-OFDM modulator output $Y[i]$ (see Fig.~\ref{fig6}). From the definition of DFZT in (\ref{seqn12988}), for a given $l \in \{0, 1, \cdots, N-1 \}$, 
\begin{eqnarray}
\label{eqncomputeyddkl}
    y_{dd}[k,l] & = & \frac{1}{\sqrt{M}} \sum\limits_{p=0}^{M-1} \left( Y[l+ pN] e^{j 2 \pi \frac{l k}{MN}} \right) \, e^{j 2 \pi p \frac{k}{M}}.
\end{eqnarray}Since $\left( Y[l+ pN] e^{j 2 \pi \frac{l k}{MN}} \right)$ depends on $k$, a direct $M$-point Inverse FFT cannot be used to compute $y_{dd}[k,l]$ for all $k=0,1,\cdots, M-1$. On the other hand, a naive implementation of (\ref{eqncomputeyddkl}) would result in a ${\mathcal O}(M)$ complexity to compute $y_{dd}[k,l]$ for each $(k,l)$ resulting in a total complexity of ${\mathcal O}(M^2N)$ for all $k=0,1,\cdots, M-1$, $l=0,1,\cdots, N-1$.

However, we can achieve a lower order of complexity since DFZT is a composition of IDFT (which converts $Y[i]$ back to discrete-time symbols $y[n]$) followed by
DZT (which converts $y[n]$ to $y_{dd}[k,l]$). The complexity of $MN$-point IDFT is ${\mathcal O}(MN \log(MN))$ and that of DZT is ${\mathcal O}(MN \log(N))$, i.e., an overall complexity of ${\mathcal O}(MN \log(MN))$.

\begin{figure}[h]
\vspace{-1mm}
\centering
\includegraphics[width=8.8cm, height=6.9cm]{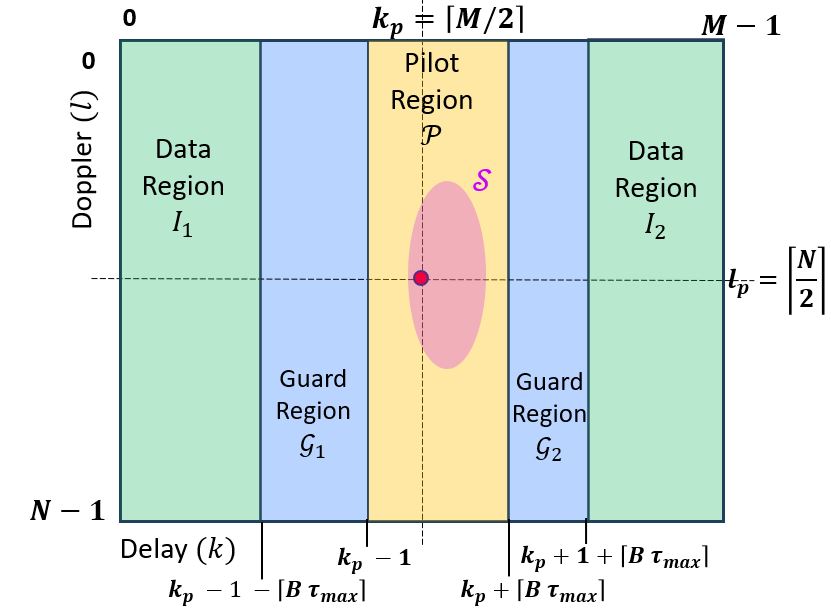}
\vspace{-2mm}
\caption{
Pilot, guard and data carriers in a $M \times N$ Zak-OTFS DD frame.}
\label{figpilotzak}
\vspace{-2mm}
\end{figure}
\section{I/O relation of the proposed Zak-OTFS over CP-OFDM and its acquisition}
\label{subseciorel}
From Theorem \ref{thm1} and Theorem \ref{thm2} we know the equivalence of the CP-OFDM based Zak-OTFS modulator/de-modulator (top chains in Fig.~\ref{fig5} and Fig.~\ref{fig6})
to that of DD domain pulse shaping and Zak transform based Zak-OTFS modulator/de-modulator (bottom chains in Fig.~\ref{fig5} and Fig.~\ref{fig6}). Due to this equivalence, the DD domain I/O
relation between $x_{dd}[k,l]$ and $y_{dd}[k,l]$ for the CP-OFDM based architecture is same as that of the Zak transform based architecture.

Therefore the I/O relation for the proposed Zak-OTFS over CP-OFDM is given by (\ref{eqniorel1}  (see also (\ref{eqniorel2}) and (\ref{eqniorel3})), with the transmit pulse shaping filter $w_{tx}(\tau, \nu)$ and the receive matched filter $w_{rx}(\tau, \nu)$ constrained to (\ref{eqnwtxzakoverofdm}) and (\ref{eqnwrxzakoverofdm})
respectively. We denote the discrete DD support set of $h_{dd}[k,l]$ by ${\mathcal H}$, i.e., the energy of $h_{dd}[k,l]$ in all taps outside ${\mathcal H}$ is significantly smaller than for those inside ${\mathcal H}$, i.e., $\sum\limits_{(k,l) \notin {\mathcal H}} \vert h_{dd}[k,l] \vert^2 \ll \sum\limits_{(k,l) \in {\mathcal H}} \vert h_{dd}[k,l] \vert^2 $.

In Fig.~\ref{figpilotzak}, we show a typical Zak-OTFS frame in the DD domain (i.e., $x_{dd}[k,l]$, $k=0,1,\cdots, M-1$, $l=0,1,\cdots, N-1$).
For the acquisition of $h_{dd}[k,l]$ a point pilot is transmitted in the middle of the frame at $(k,l) = (k_p, l_p) = (\lceil M/2 \rceil, \lceil N/2 \rceil)$. Due to channel delay and Doppler spread, the received pilot is no more localized at $(k_p, l_p)$ but gets smeared over the taps $(k_p, l_p) + {\mathcal H}$ which we denote by ${\mathcal S}$ (see Fig.~\ref{figpilotzak}).
The pilot region ${\mathcal P}$ is dedicated for the acquisition of $h_{dd}[k,l]$ and is designed to contain ${\mathcal S}$.
Let $\tau_{max} \Define \max_i \tau_i$ is the maximum path delay (the minimum path delay is zero). Since the delay domain resolution is $1/B$ (see information lattice in (\ref{lambdap})), the number of discrete delay bins corresponding to the channel delay spread is approximately $B \tau_{max}$. We consider ${\mathcal P} = \{ (k,l) \, | \, k_p - 1 \leq k \leq k_p + \lceil B \tau_{max} \rceil, l =0,1,\cdots, N-1\}$.  The pilot region is surrounded by the guard regions ${\mathcal G}_1= \{ (k,l) \, \vert \, k_p - 1 - \lceil B \tau_{max} \rceil \, \leq k \leq k_p - 2 \,,\, l = 0,1,\cdots, N-1 \}$ and
${\mathcal G}_2 = \{ (k,l) \, \vert \, k = k_p + 1 + \lceil B \tau_{max} \rceil \,,\, l = 0,1,\cdots, N-1 \} $ so as to minimize the interference between data and pilot carriers. Note that data/information symbols are
carried by carriers at DD taps which are not in ${\mathcal G}_1 \, \cup \, {\mathcal P} \, \cup \, {\mathcal G}_2$. The pilot and guard region overhead is therefore $N(2 \lceil B \tau_{max} \rceil + 3)$ out of the total $MN$ DD carriers.
The pilot overhead fraction is therefore
\begin{eqnarray}
\label{pilotfrac}
    \frac{N (2 \lceil B \tau_{max} \rceil + 3)} {MN} = \frac{(2 \lceil B \tau_{max} \rceil + 3)}{M}.
    \end{eqnarray}
Note that the pilot is designed as a strip along the Doppler axis
for scenarios where we expect a fixed delay spread but where the Doppler spread can vary from medium to high.
Therefore, $\tau_p$ must satisfy
\begin{eqnarray}
\label{taupcnd}
   B \tau_p & > &   (2 \lceil B \tau_{max} \rceil + 3).
\end{eqnarray}
Let $\nu_{max}$ denote the
maximum Doppler shift of any channel path.
Then, the received pilot (with support set ${\mathcal S}$) has a spread of approximately $\lceil 2 T \nu_{max} \rceil$ along the discrete Doppler axis and therefore for reliable and accurate acquisition of $h_{dd}[k,l]$ from the received pilot, the crystallization condition needs to be satisfied, i.e., the discrete Doppler period $N = T \nu_p$ must be greater than $ 2 T \nu_{max} $
\begin{eqnarray}
\label{cryscnd}
    T \nu_p & > &  2 T \nu_{max}.
\end{eqnarray}From the received pilot $y_{dd}[k,l]$, $(k,l) \in {\mathcal S}$,
an estimate of $h_{dd}[k,l]$, $(k,l) \in {\mathcal H}$ is given by the cross-ambiguity between the received pilot and the transmitted point pilot $x_{p,dd}[k,l] = \sum\limits_{n,m \in {\mathbb Z}} e^{j 2 \pi \frac{n l}{N}} \delta[k - k_p - nM] \delta[l - l_p - mN]$ (see ($22$) in \cite{spreadpaper} for more details), i.e. for all $(k,l) \in {\mathcal H}$, an estimate of $h_{dd}[k,l], (k,l) \in {\mathcal H}$ is given by
\begin{eqnarray}
    {\widehat h}_{dd}[k,l] & \hspace{-3mm} = & \hspace{-3mm} \sum\limits_{(k',l') \in {\mathcal S}} \hspace{-3mm} y_{dd}[k',l'] \, x_{p,dd}^*[k' - k , l' - l] \, e^{-j 2 \pi \frac{l(k' - k)}{MN}}, \nonumber \\
    x_{p,dd}[k,l] & \hspace{-3mm} \Define & \hspace{-3mm} \sum\limits_{n,m \in {\mathbb Z}} e^{j 2 \pi \frac{n l}{N}} \delta[k - k_p - nM] \delta[l - l_p - mN].
\end{eqnarray}

\section{Salient features and advantages of the proposed Zak-OTFS over CP-OFDM transceiver architecture}
\label{subsecsalient}

\subsection{Benefits of Zak-OTFS modulation can be achieved with existing CP-OFDM based standardized modems}
The proposed Zak-OTFS over CP-OFDM transceiver architecture discussed in
Section \ref{zakoverofdmtx} and Section \ref{zakoverofdmrx} allows Zak-OTFS to be implemented in existing CP-OFDM based standardized modems since the precoding (with IDFZT) at the transmitter and post-processing (DFZT) at the receiver does not require additional hardware and can be performed at low complexity (only ${\mathcal O}(K \log K)$, $K=MN=BT$ is the number of carriers) through software implementation. This enables us to achieve benefits of Zak-OTFS modulation (robustness to channel delay and Doppler spread, low pilot overhead, almost a stationary I/O relation in DD domain) on existing wireless infrastructure. 

\subsection{CP-OFDM is Zak-OTFS over CP-OFDM with minimum delay period}
In the proposed transceiver architecture (top chains in Fig.~\ref{fig5} and Fig.~\ref{fig6}), we can choose the delay and Doppler periods $M$ and $N$ arbitrarily as long as their product $MN$ equals the number of carriers $BT$. The special case of Zak-OTFS with $M=1$ and $N = BT$, and transmit pulse shaping filter and receive matched filter as in (\ref{eqnwtxzakoverofdm}) and
(\ref{eqnwrxzakoverofdm}) respectively, is nothing but CP-OFDM.
This is not difficult to see, since with $M=1$ the IDFZT precoding in (\ref{sieqn12}) reduces to
\begin{eqnarray}
    S[i] & = & x_{dd}[0, i], \,\,\, \text{when} \,\, M =1,
\end{eqnarray}i.e., the symbol transmitted on the $i$-th CP-OFDM sub-carrier is simply the DD domain symbol $x_{dd}[0, i]$ on the $i$-th Doppler bin. Similarly, at the receiver, the DFZT post-processing in (\ref{seqn12988}) reduces to
\begin{eqnarray}
y_{dd}[0,l] & = & Y[l], \,\,\, \text{when} \,\, M =1,
\end{eqnarray}i.e., the symbol received in the $l$-th Doppler bin is same as the FD symbol received on the $l$-th CP-OFDM sub-carrier. Therefore, switching between Zak-OTFS over CP-OFDM and CP-OFDM can be carried out in software by choosing $M=1$. Note that
$\tau_p = M/B$, and therefore $M=1$ corresponds to the minimum possible delay period. 

\subsection{Zak-OTFS over CP-OFDM as a family of modulations}
Since $MN = BT$, and $M$ and $N$ are positive integers,
the proposed Zak-OTFS over CP-OFDM is in fact a family of modulations, wherein each possible pair of positive integers $(M,N)$ ($MN = BT$) corresponds to a different modulation. The two extreme special cases are $M=1$, which as we have seen corresponds to CP-OFDM itself, and the other extreme is
$N=1$ which corresponds to the
the minimum possible Doppler period of $\nu_p = 1/T$. In fact, simulations for high Doppler spread scenarios considered in Section \ref{numsec} reveal that Zak-OTFS over CP-OFDM achieves highest effective SE when $N=1$. Varying $(M,N)$
allows to resize the delay and Doppler periods so as to ensure that the conditions (\ref{taupcnd}) and (\ref{cryscnd}) are satisfied (which ensure robustness to channel delay and Doppler spread). Resizing $(M,N)$ also helps to optimize the pilot overhead.
Compared to CP-OFDM $(M=1$), Zak-OTFS over CP-OFDM therefore provides
an extra degree of freedom in choosing $(M,N)$
so as to optimize performance for channel scenarios with
different delay and Doppler profiles.

\subsection{Filtered-OFDM as a special case of Zak-OTFS over CP-OFDM}
Filtered-OFDM differs from CP-OFDM in terms of the time windows $W_2(t)$ and $W_4(t)$. In the proposed transceiver architecture (see Fig.~\ref{fig5} for transmitter and Fig.~\ref{fig6} for receiver), with $M=1$ and appropriately chosen time-windows we can implement filtered-OFDM.

\section{{Numerical results}}
\label{numsec}
In the following we compare the effective spectral efficiency (SE) achieved by CP-OFDM, proposed Zak-OTFS over CP-OFDM and unconstrained Zak-OTFS (where we are free to choose the period parameters and the pulse shaping and matched filter).
For Zak-OTFS over CP-OFDM and for CP-OFDM we consider the standardized 3GPP 5G NR numerology \cite{3GPP}.

We consider communication in Frequency Range 3 (FR3) band with carrier frequency $f_c = 15, 7$ GHz, and also Frequency Range 1 (FR1).  Possible standardized sub-carrier spacings (SCS) for CP-OFDM are $\Delta f = 15, 30, 60$ KHz. We simulate four scenarios as listed in Table \ref{tab3}.
Scenarios-I and II model non-line-of-sight (NLOS) communication with a user in a vehicle (car) and train respectively. Scenario-III models LOS aircraft-to-ground (A2G) communication with an aircraft speed of $1000$ km/hr (IMT-2030 considers mobile speeds up to $1000$ Km/hr). Scenario IV models NLOS for a long delay profile with no mobility $\nu_{max} = 0$ Hz, $f_c = 3.5$ GHz. 

    \begin{table}[h!]
		\caption{{Simulation scenarios}}
		\label{tab3}
		\centering
		\begin{tabular}
			{ | c|| c| c| c| c|c| }
			\hline
			\multicolumn{6}{|c|}{Simulation scenarios} \\
			\hline
			Scenario & Carrier & Channel & Speed & Max. & Delay\\
            & Freq. & model & (km/hr) & Doppler & profile\\
            &       &       &         &   shift      & (UMa) \\
			\hline
            I & $15$ GHz & TDL-C & $90$ & $1.25$ KHz & Normal  \\
			\hline
            II & $15$ GHz & TDL-C & $216$ &  $3$ KHz & Normal  \\
			\hline
            III & $7$ GHz & TDL-D & $1000$ &  $6.48$ KHz & Short  \\
			\hline
            IV & $3.5$ GHz & TDL-C & $0$ &  $0$ KHz & Long  \\
			\hline
		\end{tabular}
	\end{table} 

The tapped delay line (TDL) models and the multi-path power-delay profiles are as per 3GPP TR 38.901 \cite{TR38901}.
For the normal and short delay profiles (UMa - Urban Macro), we consider the path delay scaling factor to be $302$ ns and $85$ ns respectively (see Table 7.7.3-2 in \cite{TR38901}).
The Doppler shift of the $i$-th channel path is given by $\nu_i = \nu_{max} \cos(\theta_i), i=1,2,\cdots, N_p$ ($N_p$ is the total number of paths), where $\nu_{max}$ is the maximum possible Doppler shift of any path and $\theta_i, i=1,2,\cdots, N_p$
are i.i.d. uniformly distributed in $[0 \,,\, 2 \pi)$.

For standardized 3GPP 5G NR CP-OFDM, we consider $48$ sub-carriers and $14$ symbols. For example, with $\Delta f = 15$ KHz, the total time duration is $1$ ms and bandwidth is $720$ KHz. With $\Delta f = 30$ KHz, these become $0.5$ ms and $1440$ KHz respectively. For normal delay profile, the maximum delay spread is $2.6$ ms, and therefore with $60$ KHz SCS we consider extended CP as specified in the 5G NR standard (in this case there are $12$ and not $14$ symbols in a CP-OFDM sub-frame of $0.25$ ms duration). 
We optimize the effective SE w.r.t. all possible Type-A pilot arrangements in 3GPP 5G NR, $\Delta f = 15, 30, 60$ KHz and Modulation and Coding Schemes (MCS) as specified in MCS Table index $1$ (Table 5.1.3.1-1 in \cite{TS38214}). Note that the LDPC code is spread over all $14$ symbols in the CP-OFDM sub-frame. For a given sub-carrier spacing, pilot arrangement and MCS, the effective SE is given by $(1 - \text{BLER})$ times  the total number of information bits (before LDPC coding) divided by the total time-bandwidth product (for example, for $\Delta f = 15$ KHz, the total time-bandwidth product is $720$ KHz $\times 1$ ms which is $720$). \text{BLER} denotes the average block error rate, and in our optimization we only consider those operating points (MCS, $\Delta f$ and pilot arrangement) for which $\text{BLER} < 0.1$. 

For the proposed Zak-OTFS over CP-OFDM, since $\Delta f = 15, 30, 60$ Hz, the variable $T$ for Zak-OTFS over CP-OFDM is constrained to be $T = 66.66, 33.33, 16.66 \mu s$. These correspond to Doppler resolution of $1/T = 15, 30, 60$ KHz. For scenario-I, $\nu_{max} = 1.25$ KHz, for which $N = T \nu_p$ must be greater than $ T (2 \nu_{max})$ for (\ref{cryscnd}) to be satisfied.
For $T = 66.66, 33.33, 16.66 \mu s$, the corresponding values of $ T (2 \nu_{max})$ are $1/6, 1/12, 1/24$. Since all these three values are less than $1$, and $N$ is a positive integer, the smallest possible value of $N$ is $1$.
Also, $M = B \tau_p$ must be greater than the total width $ (2 \lceil B \tau_{max} \rceil + 3)$ of the pilot and guard region (see (\ref{taupcnd})). The pilot overhead fraction is therefore $N(2 \lceil B \tau_{max} \rceil + 3)/(BT)$ (see (\ref{pilotfrac})).
For all Zak-OTFS over CP-OFDM numerologies, $BT = 48$, and therefore, for $T = 66.66, 33.33, 16.66 \mu s$, the corresponding values of the pilot overhead fraction are $7N/48$, $11N/48$ and $19N/48$ ($\tau_{max}$ is $2.6 \, \mu s$). In other words for a given $N$, the pilot overhead fraction increases linearly with decreasing $T$.
For the smallest possible overhead while still satisfying the conditions in (\ref{taupcnd}) and (\ref{cryscnd}), we therefore choose $N=1$ and $T = 66.66 \, \mu s$, and therefore $\nu_p = 15$ KHz, $\tau_p = 66.66 \, \mu s$. Note that this corresponds to Zak-OTFS over CP-OFDM with minimum Doppler period. The same is also true for scenario-II where $\nu_{max} = 3$ KHz.

For scenario-III, $\nu_{max} = 6.48$ KHz, and for the short delay profile $\tau_{max} = 1.17 \, \mu s$. As per the condition in (\ref{cryscnd}), the smallest possible value of $N$ is still $1$.
For $T = 66.66, 33.33, 16.66 \mu s$, the corresponding values of the pilot overhead fraction are $5N/48$, $7N/48$ and $11N/48$. With $T=66.66 \, \mu s$,
we achieve the lowest pilot overhead.
However, with $T=66.66 \, \mu s$ and $N=1$, $\nu_p = 15$ KHz, the crystallization condition in (\ref{cryscnd}) is satisfied with a very small margin (since $2 \nu_{max} \approx 13$ KHz which is close to $15$ KHz). This results in performance degradation due to some amount of Doppler domain aliasing in practical systems since the condition in (\ref{cryscnd}) does not strictly account for the shape of the Doppler domain pulse shaping filter. Therefore, for scenario-III we consider $T= 33.33 \mu s$ and $N=1$.
The Zak-OTFS over CP-OFDM numerology for the three scenarios is summarized in Table \ref{tab4}.
Note that the CP duration is equal to that for a standardized
CP-OFDM symbol having same $T$. Also, for Zak-OTFS over CP-OFDM, we optimize the effective SE w.r.t. MCS.  The LDPC code is spread over all $14$ Zak-OTFS over CP-OFDM packets. The arrangement of DD domain data and pilots in a packet/frame is as shown in Fig.~\ref{figpilotzak}. The total energy of data
symbols is equal to that of the pilot (as this has been found to result in achieving a good trade-off between pilot interference to data symbols and channel estimation accuracy). For all scenarios, detection of the transmitted DD domain symbols $x_{dd}[k,l]$ from the received DD domain symbols $y_{dd}[k,l]$ is performed
using MMSE equalization of the matrix-vector form of the DD domain I/O relation (see \cite{ZAKOTFS2, otfsbook}). The soft detected symbols from all $14$ received Zak-OTFS over CP-OFDM packets are then forwarded to the LDPC decoder.

Finally we also simulate a single Zak-OTFS packet/frame of duration $T=1$ ms with no numerology constraint and no constraint on the choice of DD domain transmit pulse shaping and matched filtering at the receiver. The packet bandwidth is $B = 720$ KHz.
For scenarios-I and II we choose a Doppler period $\nu_p=15$ KHz and delay period $\tau_p = 1/\nu_p = 66.66 \, \mu s$, which satisfies both the conditions in (\ref{taupcnd}) and (\ref{cryscnd}). For scenario-III, the channel Doppler spread is higher ($6.48$ KHz) and therefore we consider $\nu_p = 24$ KHz. For the unconstrained Zak-OTFS we have only one packet for which $MN  = B T = 720$. The arrangement of DD domain data and pilots in a packet/frame is as shown in Fig.~\ref{figpilotzak}. The total energy of data
symbols is equal to that of the pilot. For a given $\nu_p$, $N = T \nu_p$, and therefore the pilot overhead fraction is given by $N(2 \lceil B \tau_{max} \rceil + 3)/(BT) = \nu_p (2 \lceil B \tau_{max} \rceil + 3)/B$. The Zak-OTFS (unconstrained) numerology for the three scenarios is given in Table \ref{tab5}.
We consider DD domain Gauss-sinc pulse shaping filter $w_{tx}(\tau, \nu)$ as described in \cite{Arpan2025}. The receiver matched filter is given by $w_{rx}(\tau, \nu) = w_{tx}^*(-\tau, - \nu) \, e^{j 2 \pi \nu \tau}$ as it optimizes the received signal-to-noise ratio  \cite{ZAKOTFS2,otfsbook,Hanly2024}. For unconstrained Zak-OTFS also, we optimize the effective SE w.r.t. MCS. Another
important benefit with unconstrained Zak-OTFS is that only a single guard time interval (of duration equal to the maximum channel delay spread is required at the beginning of the packet so as to avoid interference from any previous packet transmission). The overhead of this guard time interval is small, for example in scenarios-I and II this overhead is $4.7 \mu s/ 1 ms$ which is only $0.47$ percent and in scenario-III it is even smaller (only $0.117$ percent). DD domain equalization is performed using MMSE equalization of the matrix-vector form of the DD domain I/O relation, followed by LDPC decoding.
   \begin{table}[h!]
		\caption{{Zak-OTFS over CP-OFDM numerology}}
		\label{tab4}
		\centering
		\begin{tabular}
			{ | c|| c| c| c| c|c| }
			\hline
			Scenario & $\frac{1}{T}$ & $B$ & $(M,N)$ & Pilot & CP\\
                     &  & (KHz) &  & overhead & ($\mu s$)\\
			\hline
            I & $15$ KHz & $720$ & $(48,1)$ & $\frac{7}{48}$ & $4.7$  \\
			\hline
            II & $15$ KHz & $720$ & $(48,1)$ &  $\frac{7}{48}$ &  $4.7$  \\
			\hline
            III & $30$ KHz & $1440$ & $(48,1)$ &  $\frac{7}{48}$ & $2.34$  \\
			\hline
		\end{tabular}
	\end{table} 

       \begin{table}[h!]
		\caption{{Zak-OTFS (unconstrained) numerology}}
		\label{tab5}
		\centering
		\begin{tabular}
			{ | c|| c| c| c| c|c| }
			\hline
			Scenario & $\frac{1}{T}$ & $B$ & $\nu_p$ & $(M,N)$ & Pilot \\
                     &               & (KHz) & (KHz) &  & overhead \\
			\hline
            I,II & $1$ KHz & $720$ & $15$ & $(48,15)$ & $\frac{7}{48}$  \\
			\hline
            III & $1$ KHz & $720$ & $24$ & $(30, 24)$ & $\frac{8}{48}$  \\
			\hline
		\end{tabular}
	\end{table} 

\begin{figure}[h]
\vspace{-1mm}
\centering
\includegraphics[width=9.5cm, height=6.7cm]{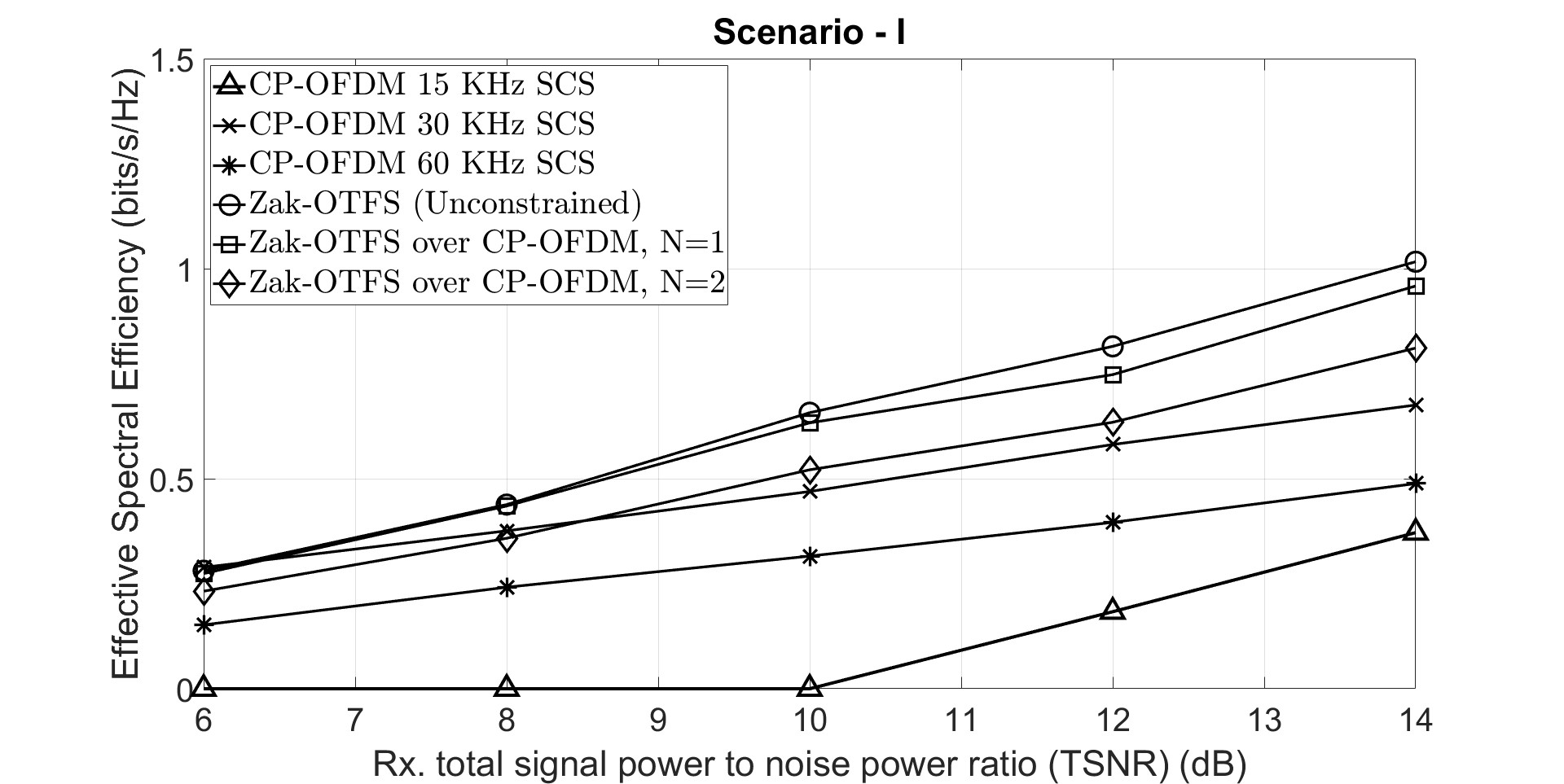}
\vspace{-2mm}
\caption{
Effective spectral efficiency vs. total received signal power to noise ratio (TSNR) (dB), for Scenario-I (see Table \ref{tab3}) and a fixed $\nu_{max}=1.25$ KHz. Total power includes both data and pilot power.}
\label{figsim1}
\vspace{-2mm}
\end{figure}
In Fig.~\ref{figsim1} we plot the effective SE as a function of increasing total received signal power (data and pilots) to noise ratio (TSNR). CP-OFDM with $30$ KHz SCS, achieves better SE than with SCS of $15$ KHz and $60$ KHz. For a SCS of $15$ KHz, channel acquisition along time domain is not accurate since
for a Doppler spread of $2 \nu_{max} = 2.5$ KHz, the coherence time is roughly $0.1$ ms and the maximum number of pilots in a sub-frame is four, i.e., one pilot every $0.25$ ms. For a SCS of $60$ KHz (with extended CP), the channel acquisition along time is good (one pilot every $0.0625$ ms), but that along frequency is not as good since the coherence bandwidth is approximately $100$ KHz whereas the maximum number of pilot sub-carriers along frequency is one in every two sub-carriers, i.e., pilots repeat every $120$ KHz. Also, due to the extended CP, the CP overhead is higher than that for SCS $30$ KHz.  

In Fig.~\ref{figsim1}, Zak-OTFS over CP-OFDM ($N=1$) achieves better SE than CP-OFDM (SCS $30$ KHz) for TSNR greater than $6$ dB. At a TSNR of $14$ dB, the proposed Zak-OTFS over CP-OFDM ($N=1$) achieves roughly $40$ percent higher SE than CP-OFDM (SCS $30$ KHz). This shows that even when we constrain Zak-OTFS with CP-OFDM numerology and filters (pulse shaping and matched filtering), it still achieves significant SE improvement.
We note that Zak-OTFS over CP-OFDM with $N=2$ has an inferior SE performance when compared to that with $N=1$ since the pilot overhead fraction is $14/48$ with $N=2$ versus $7/48$ with $N=1$, i.e., a difference of around $15$ percent.

In Fig.~\ref{figsim1} we also note that for TSNR greater than $8$ dB, unconstrained Zak-OTFS achieves even better SE than Zak-OTFS over CP-OFDM.
This is primarily due to the very low guard-time overhead between two consecutive transmissions in unconstrained Zak-OTFS (only $0.47$ percent in Zak-OTFS compared to roughly $6.5$ percent in Zak-OTFS over CP-OFDM).

\begin{figure}[h]
\vspace{-1mm}
\centering
\includegraphics[width=9.5cm, height=6.7cm]{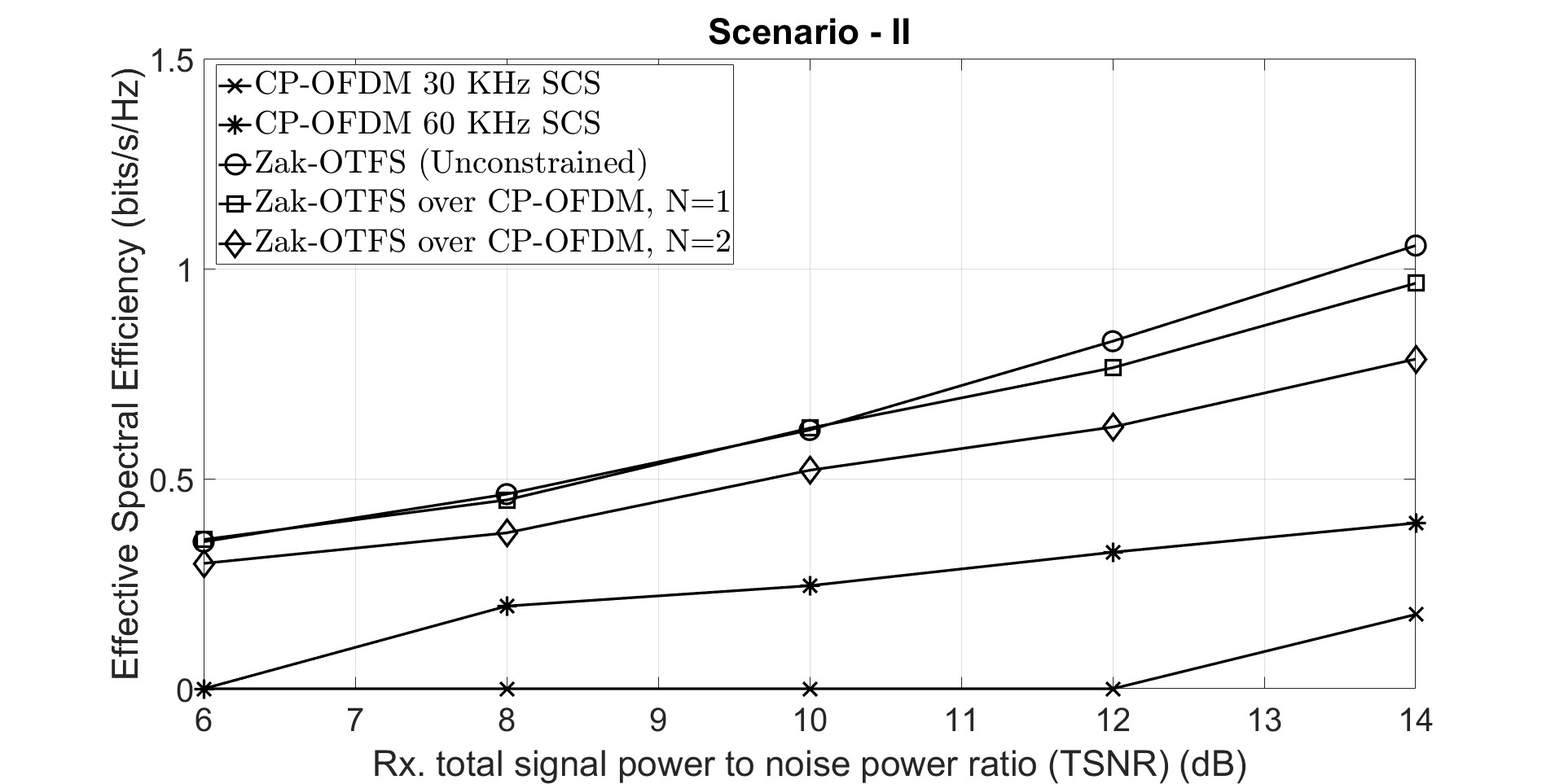}
\vspace{-2mm}
\caption{
Effective spectral efficiency vs. total received signal power to noise ratio (TSNR) (dB), for Scenario-II (see Table \ref{tab3}) and a fixed $\nu_{max}=3$ KHz. Total power includes both data and pilot power.}
\label{figsim2}
\vspace{-2mm}
\end{figure}
In Fig.~\ref{figsim2}, we plot the SE for scenario-II for $\nu_{max} = 3$ KHz. We observe that, for CP-OFDM the highest SE is achieved with $60$ KHz SCS since the maximum Doppler spread is now $6$ KHz as compared to $2.5$ KHz in scenario-I. At a TSNR of $14$ dB, Zak-OTFS over CP-OFDM ($N=1$) achieves an SE improvement of almost $145$ percent over CP-OFDM. For TSNR greater than $10$ dB, unconstrained Zak-OTFS achieves an SE even better than that of Zak-OTFS over CP-OFDM.

\begin{figure}[h]
\vspace{-1mm}
\centering
\includegraphics[width=9.5cm, height=6.7cm]{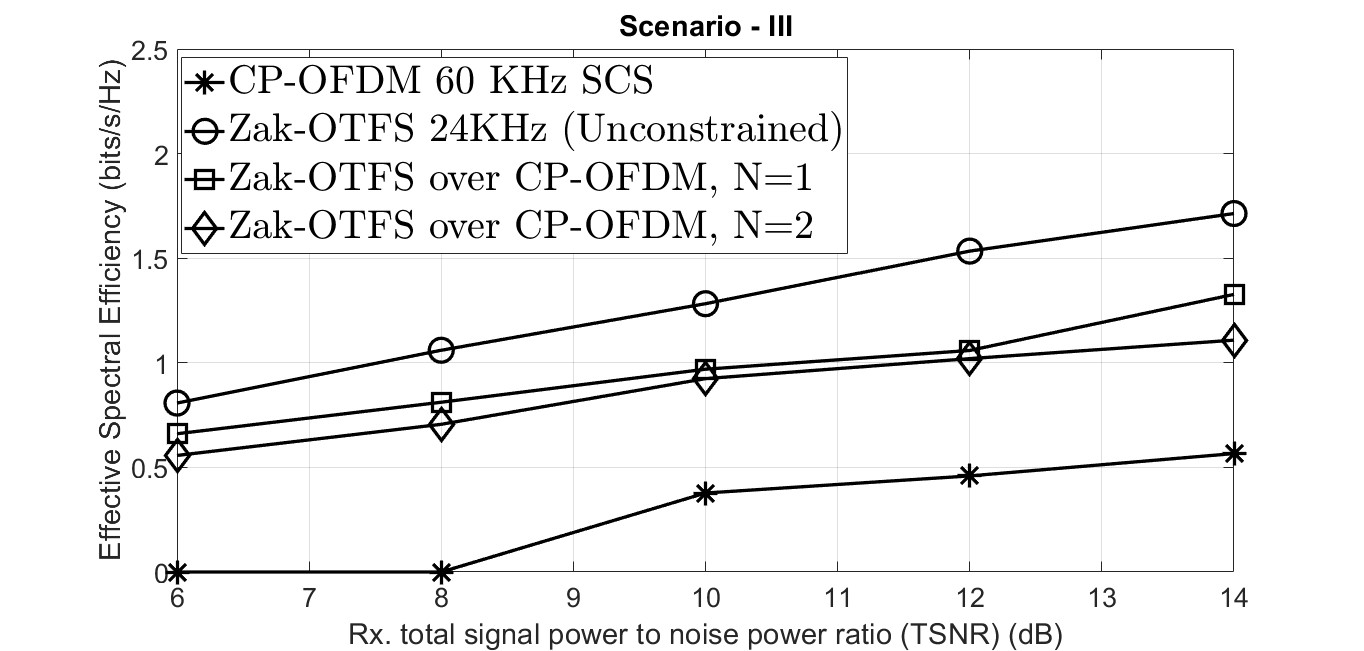}
\vspace{-2mm}
\caption{
Effective spectral efficiency vs. total received signal power to noise ratio (TSNR) (dB), for Scenario-III (see Table \ref{tab3}) and a fixed $\nu_{max}=6.48$ KHz. }
\label{figsim3}
\vspace{-2mm}
\end{figure}
In Fig.~\ref{figsim3}, we plot the SE for scenario-III for a high $\nu_{max} = 6.48$ KHz. For CP-OFDM, the best SE performance is achieved with $60$ KHz SCS, although it is zero for TSNR less than or equal to $8$ dB. This is because for low TSNR, CP-OFDM with $60$ KHz SCS is unable to achieve a BLER below $0.1$ for any MCS. We also observe that for TSNR $=14$ dB, Zak-OTFS over CP-OFDM ($N=1$) 
achieves SE which is $150$ percent more than that achieved by CP-OFDM with $60$ KHz SCS. Note that this gain is even higher than the $145$ percent gain in scenario-II and the $40$ percent gain in scenario-I. This shows that when compared to CP-OFDM, the SE improvement achieved by the proposed Zak-OTFS over CP-OFDM increases with increasing channel Doppler spread. 

Another interesting observation in Fig.~\ref{figsim3} is that
for a TSNR of $14$ dB, unconstrained Zak-OTFS ($\nu_p = 24$ KHz) achieves an SE which is
roughly $30$ percent more than that achieved by Zak-OTFS over CP-OFDM with $N=1$. This is due to the freedom in choosing the filters and the Zak-OTFS period parameters. This shows the potential of achieving significant improvements in SE by not constraining Zak-OTFS numerology and filters.

\begin{figure}[h]
\vspace{-1mm}
\centering
\includegraphics[width=9.5cm, height=6.5cm]{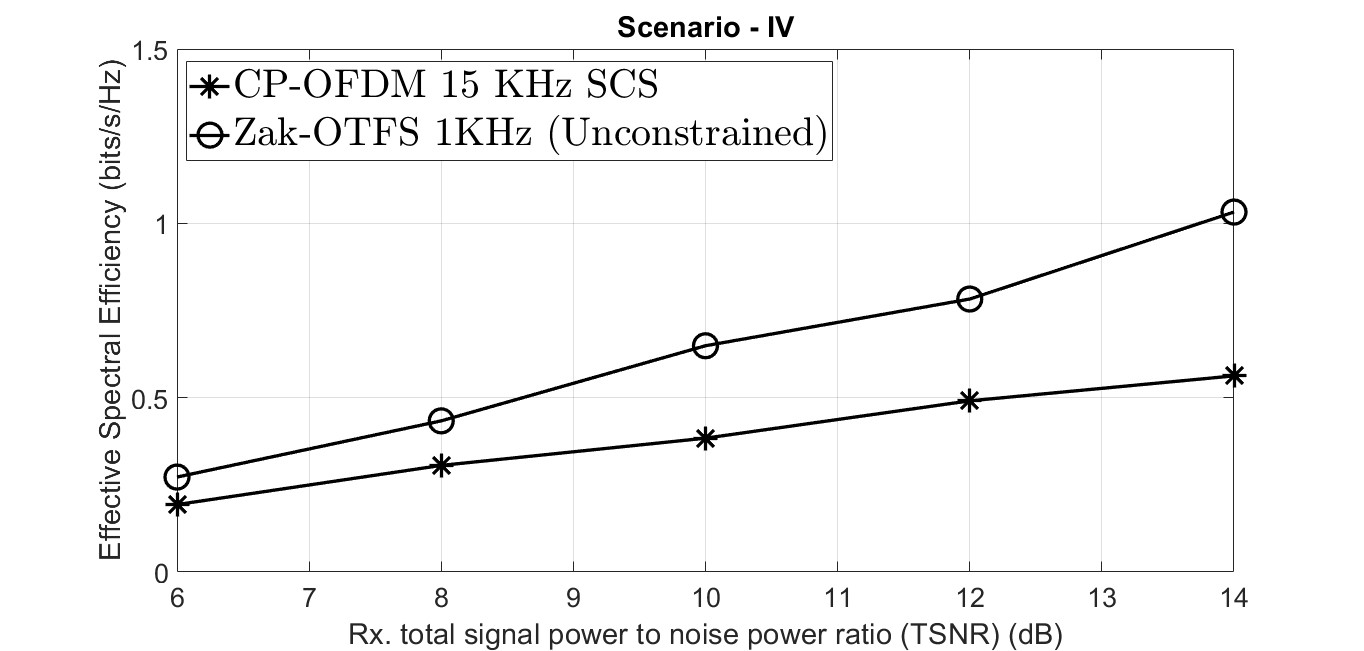}
\vspace{-2mm}
\caption{
Effective spectral efficiency vs. total received signal power to noise ratio (TSNR) (dB), for Scenario-IV (no mobility, long delay profile) (see Table \ref{tab3}).  }
\label{figsim4}
\vspace{-2mm}
\end{figure}
In Fig.~\ref{figsim4} we plot the effective SE for Scenario-IV (TDL-C, long delay profile in Table 7.7.3-2 of \cite{TR38901}, no mobility). The maximum delay spread is $9.93 \, \mu s$. Since, there is no mobility,
for CP-OFDM it suffices to choose a $15$ KHz SCS. For unconstrained Zak-OTFS it suffices to choose the minimum possible Doppler period which satisfies the conditions in (\ref{taupcnd}) and (\ref{cryscnd}). Since $\nu_p = N/T$ and $T = 1$ ms, the minimum possible Doppler period which satisfies (\ref{taupcnd}) and (\ref{cryscnd}) is $\nu_p = 1/T = 1$ KHz which corresponds to $N=1$ and $M = 720$. Since the unconstrained Zak-OTFS pilot overhead fraction is $19N/720$, $N=1$ also corresponds to minimum pilot overhead.

In Fig.~\ref{figsim4} it is observed that the effective SE performance of CP-OFDM ($15$ KHz SCS) is significantly inferior to that of unconstrained Zak-OTFS (with $\nu_p = 1$ KHz). This is due to the degradation in CP-OFDM resulting from inter-symbol interference (ISI) as the CP duration of $4.7 \mu s$ is less than the delay spread of $9.93 \mu s$.
Note that CP-OFDM is nothing but Zak-OTFS over CP-OFDM
with minimum delay period ($M=1$). For this no mobility scenario, we observed that the highest SE was achieved with $M=1, N=48$ and so we have not plotted the curves for other $(M,N)$ configurations.

\appendices

\section{Proof of Result \ref{res01}}
\label{prfres01}
Since $s(t)$ in (\ref{steqn}) is band-limited to the FD interval $\left[ - \frac{\Delta f}{2} \,,\, K \Delta f - \frac{\Delta f}{2} \right]$
having bandwidth $B = K \Delta f$, from the sampling theorem \cite{signalsystemoppenheim} it follows that $s(t)$ can be reconstructed from its samples taken at the sampling rate $B$, i.e.,
\begin{eqnarray}
\label{steqn86347}
    s(t) & = &  \frac{1}{\sqrt{K \Delta f}} \, w_1(t) \, \star \, \left( \sum\limits_{n \in {\mathbb Z}} s[n] \, \delta\left( t - \frac{n}{B} \right) \right),
\end{eqnarray}where the samples $s[n]$ are given by
\begin{eqnarray}
\label{stq9736}
    s[n] & = & s\left(t = \frac{n}{B} \right) \nonumber \\
    & \mya & \sum\limits_{i=0}^{K-1} S[i] \, e^{j 2 \pi n \frac{i}{K}}  \,  =  \,  \sqrt{K} \, x[n]
\end{eqnarray}where step (a) follows from the expression of $s(t)$
in (\ref{steqn}). Here we have also used the fact that $B T = K \Delta f T = K$. The last step follows from (\ref{xneqn19834}).
Using (\ref{stq9736}) in (\ref{steqn86347}) then gives
\begin{eqnarray}
    s(t) & = &  \frac{1}{\sqrt{K \Delta f}} w_1(t) \, \star \, \left( \sqrt{K} \sum\limits_{n \in {\mathbb Z}} x[n] \, \delta\left( t - \frac{n}{B} \right) \right) \nonumber \\
    & = & w_1(t) \, \star \, \underbrace{\left( \sqrt{T} \sum\limits_{n \in {\mathbb Z}} x[n] \, \delta\left( t - \frac{n}{B} \right) \right)}_{= {\Tilde x}(t)},
\end{eqnarray}which is exactly the IDFT based signal processing in the top chain in Fig.~\ref{fig0}.

\section{Proof of Result \ref{res1}}
\label{prfres1}
We show the equivalence in two steps.
Firstly we show that $x_{dd}(\tau, \nu)$ in (\ref{eqn3494}) is the Zak transform, i.e., DD realization of ${\Tilde x}(t)$, and secondly we show that the DD domain transmit pulse shaping in (\ref{pulseshapingtx}) with $w_{tx}(\tau, \nu)$ in (\ref{wtxeqn1937}) is equivalent to the linear convolution followed by time-windowing in (\ref{idztxeqn}). For the first step, using (\ref{eqnxtzak}), the inverse Zak transform of $x_{dd}(\tau, \nu)$ is given by

{\vspace{-4mm}
\small
\begin{eqnarray}
\label{eqn29267}
    {\mathcal Z}_t^{-1}\left( x_{dd}(\tau, \nu)\right) & = & \sqrt{\tau_p} \int\limits_{0}^{\nu_p} x_{dd}(t, \nu) \, d\nu \nonumber \\
    & & \hspace{-24mm} \mya \sqrt{\tau_p} \sum\limits_{k \in {\mathbb Z}} \sum\limits_{l \in {\mathbb Z}} x_{dd}[k,l]  \delta(t - k/B) \int\limits_{0}^{\nu_p} \, \delta\left(\nu - l \frac{\nu_p}{N} \right)  d\nu \nonumber \\
    & & \hspace{-24mm} \myb \sqrt{\tau_p} \sum\limits_{k \in {\mathbb Z}} \underbrace{\sum\limits_{l=0}^{N-1} x_{dd}[k,l] }_{= \sqrt{N} x[k] } \delta(t - k/B)  \, = \, {\Tilde x}(t),
\end{eqnarray}\normalsize}where steps (a) and (b) follow from (\ref{eqn3494}) and (\ref{idzteqn92367}) respectively, and also that $T = N \tau_p$.
For the second step of the proof, note that filtering followed by time-windowing in (\ref{idztxeqn}) can be written as

{\vspace{-4mm}
\small
\begin{eqnarray}
\label{xteqn15}
    W_2(t) \, \left[  w_1(t) \, \star \, {\Tilde x}(t)\right] & & \nonumber \\
    & & \hspace{-15mm} = W_2(t) \int w_1(\tau) {\Tilde x}(t - \tau) \, d\tau \nonumber \\
    & & \hspace{-15mm} \mya \iint w_1(\tau) w_2(\nu) {\Tilde x}(t - \tau) \, e^{j 2 \pi \nu t} \, d\tau \, d\nu  \nonumber \\
   & & \hspace{-15mm} = 
    \iint w_{tx}(\tau, \nu) {\Tilde x}(t - \tau) \, e^{j 2 \pi \nu (t - \tau)} \, d\tau \, d\nu,
\end{eqnarray}\normalsize}where step (a) follows from the fact that $W_2(t) = \int w_2(\nu) \, e^{j 2 \pi \nu t} \, d\nu$ (see (\ref{wtxeqn1937})), and the next step follows from the expression of $w_{tx}(\tau, \nu)$ in (\ref{wtxeqn1937}). To complete the proof we just need to show that the Zak transform of the IDZT based transmit signal $W_2(t) \, \left[  w_1(t) \, \star \, {\Tilde x}(t)\right]$ is identical to the Zak transform of the transmit signal based on continuous Inverse Zak transform.
In (\ref{eqnxtzak}), the Zak transform of the transmit signal based on continuous Inverse Zak transform is $x_{dd}^{w_{tx}}(\tau, \nu) = w_{tx}(\tau, \nu) *_{\sigma} x_{dd}(\tau, \nu)$ and therefore it suffices to show that the Zak transform of $W_2(t) \, \left[  w_1(t) \, \star \, {\Tilde x}(t)\right]$ is $w_{tx}(\tau, \nu) *_{\sigma} x_{dd}(\tau, \nu)$. Indeed
\begin{eqnarray}
\label{eqn36413}
    {\mathcal Z}_t\left( W_2(t) \, \left[  w_1(t) \, \star \, {\Tilde x}(t)\right] \right) & & \nonumber \\
    & & \hspace{-35mm} \mya {\mathcal Z}_t\left( \iint w_{tx}(\tau', \nu') {\Tilde x}(t - \tau') \, e^{j 2 \pi \nu' (t - \tau')} \, d\tau' \, d\nu' \right) \nonumber \\
    & & \hspace{-35mm} \myb  \iint w_{tx}(\tau', \nu') {\mathcal Z}_t\left( {\Tilde x}(t - \tau') \, e^{j 2 \pi \nu (t - \tau')} \right) \, d\tau' \, d\nu'  \nonumber \\
\end{eqnarray}where step (a) follows from (\ref{xteqn15}) (with $(\tau, \nu)$ substituted with $(\tau', \nu')$) and step (b) follows from the linearity of Zak transform as an operator.
From \cite{SKOTFS2}, \cite{otfsbook}, we know that for any TD signal $a(t)$ having Zak transform $a_{dd}(\tau, \nu)$, the Zak transform of $a(t - \tau') \, e^{j 2 \pi \nu'(t - \tau')}$ is
$e^{j 2 \pi \nu' (\tau - \tau')} \, a_{dd}(\tau - \tau', \nu - \nu')$.
Using this result with $a(t) = {\Tilde x}(t)$ and using the fact that the Zak transform of ${\Tilde x}(t)$ is $x_{dd}(\tau, \nu)$ (see (\ref{eqn29267})), it follows that
the Zak transform of ${\Tilde x}(t - \tau') \, e^{j 2 \pi \nu'(t - \tau')}$ is $e^{j 2 \pi \nu' (\tau - \tau')} \, x_{dd}(\tau - \tau', \nu - \nu')$.
Using this in (\ref{eqn36413}) gives
\begin{eqnarray}
     {\mathcal Z}_t\left( W_2(t) \, \left[  w_1(t) \, \star \, {\Tilde x}(t)\right] \right) & & \nonumber \\
    & & \hspace{-46mm} =  \hspace{-1mm} \iint \hspace{-1mm} w_{tx}(\tau', \nu') \, x_{dd}(\tau - \tau', \nu - \nu') \, e^{j 2 \pi \nu' (\tau - \tau')} \, d\tau' d\nu' \nonumber \\
     & & \hspace{-46mm} = w_{tx}(\tau, \nu) \, *_{\sigma} \, x_{dd}(\tau, \nu),
\end{eqnarray}where the last step follows from the definition of the twisted convolution operation and which completes the proof.

\section{Proof of Result \ref{res31}}
\label{prfres31}
The proof is in two steps. Firstly, we show that the Zak transform of $y(t)$ in the DZT based demodulator (with time-window $W_4(t)$ in (\ref{w4teqn1})) is identical to $y_{dd}(\tau, \nu)$ (see (\ref{yddeqn183})) with $w_{rx}(\tau, \nu)$ given by (\ref{eqn826547}) (see also Fig.~\ref{fig2}).
Secondly we show that sampling $y_{dd}(\tau, \nu)$ on the information grid $\Lambda_p$ is equivalent to the DZT of the discrete-time signal obtained after sampling $y(t)$ at $B$ Hz and periodizing it with period $MN$.

Firstly, $y(t)$ in (\ref{yteqn2864}) can be written as
\begin{eqnarray}
    y(t) & = & \iint w_4(\nu') w_3(\tau') r(t - \tau') \, e^{j 2 \pi \nu' t} \, d\tau' \, d\nu' \nonumber \\
    &  & \hspace{-10mm} = \iint \underbrace{w_4(\nu') w_3(\tau') \, e^{j 2 \pi \nu' \tau'}}_{= w_{rx}(\tau', \nu') \, \mbox{\small{in}} \, (\ref{eqn826547})} r(t - \tau') \, e^{j 2 \pi \nu' (t - \tau')} \, d\tau' \, d\nu' \nonumber \\
    &  & \hspace{-10mm} = \iint w_{rx}(\tau', \nu') r(t - \tau') \, e^{j 2 \pi \nu' (t - \tau')} \, d\tau' \, d\nu',
\end{eqnarray}whose Zak transform is
\begin{eqnarray}
    {\mathcal Z}_t\left( y(t) \right) &  &  \nonumber \\
    & & \hspace{-20mm} = {\mathcal Z}_t\left( \iint w_{rx}(\tau', \nu') \, r(t - \tau') \, e^{j 2 \pi \nu'(t - \tau') } d\tau' d\nu' \right) \nonumber \\
    & & \hspace{-20mm} \mya \iint w_{rx}(\tau', \nu') \underbrace{{\mathcal Z}_t\left(r(t - \tau') \, e^{j 2 \pi \nu'(t - \tau') } \right)}_{= e^{j 2\pi \nu' (\tau - \tau')} \, r_{dd}(\tau - \tau', \nu - \nu')} \, d\tau' d\nu' \nonumber \\
    & & \hspace{-20mm} = w_{rx}(\tau, \nu) \, *_{\sigma} \, r_{dd}(\tau, \nu) \, = \, y_{dd}(\tau, \nu),
\end{eqnarray}where step (a) follows from a similar result in the proof of Result \ref{res1} (see paragraph after (\ref{eqn36413})). The last step follows from (\ref{yddeqn183}).

For the second step of the proof, since we have just shown that $y_{dd}(\tau, \nu)$ is the Zak transform of $y(t)$, from the basic definition of Zak transform we get
\begin{eqnarray}
y_{dd}(\tau, \nu)  & = & \sqrt{\tau_p} \sum\limits_{n \in {\mathbb Z}} y(\tau + n \tau_p) \, e^{-j 2 \pi n \nu \tau_p},
\end{eqnarray}and therefore from (\ref{eqnyddkl})
\begin{eqnarray}
    y_{dd}[k,l] & = & y_{dd}\left(\tau = \frac{k}{B}, \nu = \frac{l}{T}\right) \nonumber \\
    & & \hspace{-20mm} = \sqrt{\tau_p} \sum\limits_{n \in {\mathbb Z}} y(k/B + n \tau_p) \, e^{-j 2 \pi n \frac{l}{T} \tau_p} \nonumber \\
    & & \hspace{-20mm} \mya \sqrt{\tau_p} \sum\limits_{q =0}^{N-1} \sum\limits_{p \in {\mathbb Z}} y \left( \frac{k + qM + pMN}{B} \right) \,   e^{-j 2 \pi (q + pN) \frac{l}{N} } \nonumber \\
    & & \hspace{-20mm} \myb \frac{1}{\sqrt{N}} \sum\limits_{q=0}^{N-1} \left( \sum\limits_{p \in {\mathbb Z}} {\Tilde y}[k + qM + pMN]\right) e^{-j 2 \pi \frac{q l}{N}} \nonumber \\
    & & \hspace{-20mm} \myc \frac{1}{\sqrt{N}} \sum\limits_{q=0}^{N-1} y[k + qM] \, e^{-j 2 \pi \frac{q l}{N}} \, \myd \, y_{dd}[k,l]
\end{eqnarray}where in step (a) we substitute $n = (q + pN)$, where $q \in \{ 0, 1, \cdots, N-1 \}$ and $p \in {\mathbb Z}$. Here we have also used the fact that $M = B \tau_p$ and $T = N \tau_p$. Step (b) follows from (\ref{tildeyneqn}). Steps (c) and (d) follow from (\ref{periodizeeqn}) and (\ref{dzteqn2974}) respectively.

\section{Proof of Result \ref{res4}}
\label{prfres4}
Substituting the expression of $S[i]$ from (\ref{sieqn12})
into the RHS of (\ref{idft92784}) gives
\begin{eqnarray}
    \text{IDFT}(\text{IDFZT}(x_{dd}[k,l])) &  & \nonumber \\
    & & \hspace{-28mm} = \frac{1}{\sqrt{N}} \frac{1}{M} \sum\limits_{i=0}^{MN-1} \sum\limits_{k=0}^{M-1} x_{dd}[k,i] \, e^{j 2 \pi i \frac{(n - k)}{MN}}.
\end{eqnarray}Replacing $i$ with the summation variables
$m, l$ such that $i = l + mN$, where $l = 0,1,\cdots, N-1$ and $m=0,1,\cdots, M-1$, we get 

{\vspace{-4mm}
\small
\begin{eqnarray}
\label{yert82364}
\text{IDFT}(\text{IDFZT}(x_{dd}[k,l])) &   & \nonumber \\
& & \hspace{-30mm} = \frac{1}{\sqrt{N}} \frac{1}{M} \sum\limits_{l=0}^{N-1} \sum\limits_{m=0}^{M-1} \sum\limits_{k=0}^{M-1}  x_{dd}[k,(l + mN)] \, e^{j 2 \pi (l + mN) \frac{(n - k)}{MN}} \nonumber \\
& & \hspace{-30mm} \mya \frac{1}{\sqrt{N}} \frac{1}{M} \sum\limits_{l=0}^{N-1}  \sum\limits_{k=0}^{M-1}  x_{dd}[k,l]  \, e^{j 2 \pi l \frac{(n-k)}{MN}} \underbrace{\left( \sum\limits_{m=0}^{M-1} e^{j 2 \pi m \frac{(n - k)}{M}} \right)}_{= M \delta[k  \, - \, n \, \text{mod} \,  M]} \nonumber \\
& & \hspace{-30mm} = \frac{1}{\sqrt{N}}  \sum\limits_{l=0}^{N-1}   x_{dd}[n \, \text{mod} \, M,l]  \, e^{j 2 \pi l \frac{(n \, - \, n \text{mod} M)}{MN}},
\end{eqnarray}\normalsize}where step (a) follows from the fact that $x_{dd}[k,l]$ is periodic with period $N$ along the Doppler axis.
Since $(n \, - \, n \, \text{mod} \, M)$ is an integer multiple of $M$, lets say it is $q_n M$.
Therefore, from the quasi-periodicity of $x_{dd}[k,l]$ along the Doppler axis (see (\ref{qpeqn234})) it follows that
\begin{eqnarray}
\label{eqn82674}
    x_{dd}[n \, \text{mod} \, M,l]  & = & x_{dd}[n + (n \, \text{mod} \, M \, - \,  n),l] \nonumber \\
    & & \hspace{-20mm} = x_{dd}[n,l] \, e^{j 2 \pi \frac{(n \, \text{mod} \, M \, - \,  n)}{M} \frac{l}{N} }.
\end{eqnarray}Using (\ref{eqn82674}) in the last step of (\ref{yert82364}) gives
\begin{eqnarray}
\text{IDFT}(\text{IDFZT}(x_{dd}[k,l])) & =  & \frac{1}{\sqrt{N}} \sum\limits_{l=0}^{N-1} x_{dd}[n,l] \nonumber \\
& = & \text{IDZT}(x_{dd}[k,l])),
\end{eqnarray}where the last step follows from (\ref{idzteqn9364}).
This then completes the proof.

\section{Proof of Result \ref{res5}}
\label{prfres5}
Substituting the expression of $Y[i]$ in the RHS of (\ref{dft92784})
into the RHS of (\ref{seqn12988}) we get
\begin{eqnarray}
\label{eqn2754}
    \text{DFZT}(\text{DFT}(y[n])) &  & \nonumber \\
    & & \hspace{-30mm} = \frac{1}{M} \frac{1}{\sqrt{N}} \sum\limits_{n=0}^{MN-1} \sum\limits_{p=0}^{M-1} y[n] e^{j 2 \pi (k - n) \frac{(l + pN)}{MN}}  \nonumber \\
     & & \hspace{-30mm} =  \frac{1}{\sqrt{N}} \sum\limits_{n=0}^{MN-1} \hspace{-1mm} y[n] e^{j 2 \pi l \frac{(k - n)}{MN}} \underbrace{\left( \frac{1}{M} \sum\limits_{p=0}^{M-1} e^{j 2 \pi p \frac{(k - n)}{M} } \right)}_{= 1 \, \text{if} \, (n - k) \, \text{mod} \, M = 0, 0 \, \text{otherwise}} \nonumber \\
     & & \hspace{-30mm} = \frac{1}{\sqrt{N}} \sum\limits_{\substack{n=0, \\ n \equiv k \, \text{mod} \, M}}^{MN-1} y[n] \, e^{j 2 \pi l \frac{(k - n)}{MN}}.
\end{eqnarray}Since $n$ must be congruent ($\equiv$) to $k$ modulo $M$,
$n$ must be of the type $n = qM + \, k \, \text{mod} \, M$, where $q=0,1,\cdots, N-1$. Therefore, the summation in (\ref{eqn2754}) simplifies to
\begin{eqnarray}
    \text{DFZT}(\text{DFT}(y[n])) &  & \nonumber \\
    & & \hspace{-30mm} = \frac{1}{\sqrt{N}}  \sum\limits_{q=0}^{N-1} y[k \, \text{mod} \, M + qM] \, e^{-j 2\pi \frac{q l}{N}} \, e^{j 2 \pi (k - k \, \text{mod} \, M) \frac{l}{MN}}. \nonumber \\
\end{eqnarray}Since $(k \, \text{mod} \, M - k)$ is divisible by $M$, let it be equal to $\theta M$ for some integer $\theta$. Using this fact above we get
\begin{eqnarray}
\label{eqn2954}
    \text{DFZT}(\text{DFT}(y[n])) & \hspace{-3mm}   =  & \hspace{-3mm} \frac{1}{\sqrt{N}}  \sum\limits_{q=0}^{N-1} y[k  + (q + \theta)M] \, e^{-j 2\pi \frac{(q + \theta) l}{N}} \nonumber \\
    & & \hspace{-10mm} = \frac{1}{\sqrt{N}}  \sum\limits_{q=\theta}^{N-1+\theta} y[k  + q M] \, e^{-j 2\pi \frac{q l}{N}} \nonumber \\
    &  & \hspace{-10mm} \mya \frac{1}{\sqrt{N}}  \sum\limits_{q=0}^{N-1} y[k  + q M] \, e^{-j 2\pi \frac{q l}{N}}, \nonumber \\
    &  & \hspace{-10mm} = \text{DZT}(y[n])
\end{eqnarray}where step (a) follows from the fact that $y[n]$ is periodic with period $MN$. The last step follows from (\ref{dzteqn9364}), which then completes the proof.

\end{document}